\documentclass[review,10pt,3p]{elsarticle}
\usepackage{amssymb,amsmath}
\usepackage{txfonts,multirow}
\usepackage{color,ulem}
\usepackage[colorlinks,hyperindex,dvipdfm]{hyperref}

\newcommand{\figdraft}{false} 

\journal{Elsevier}

\begin{document}

\begin{frontmatter}

\title{Third-order analysis of pseudopotential lattice Boltzmann model for
multiphase flow}

\author[SJTU]{Rongzong Huang}
\author[SJTU]{Huiying Wu \corref{Wu}}
\ead{whysrj@sjtu.edu.cn}

\cortext[Wu]{Corresponding author}
\address[SJTU]{Key Laboratory for Power Machinery and Engineering of
Ministry of Education, School of Mechanical Engineering, \\Shanghai Jiao
Tong University, Shanghai 200240, China}

\begin{abstract}
In this work, a third-order Chapman-Enskog analysis of the
multiple-relaxation-time (MRT) pseudopotential lattice Boltzmann (LB)
model for multiphase flow is performed for the first time. The leading
terms on the interaction force, consisting of an anisotropic and an
isotropic term, are successfully identified in the third-order macroscopic
equation recovered by the lattice Boltzmann equation (LBE), and then new
mathematical insights into the pseudopotential LB model are provided. For
the third-order anisotropic term, numerical tests show that it can cause
the stationary droplet to become out-of-round, which suggests the
isotropic property of the LBE needs to be seriously considered in the
pseudopotential LB model. By adopting the classical equilibrium moment or
setting the so-called ``magic'' parameter to $1/12$, the anisotropic term
can be eliminated, which is found from the present third-order analysis
and also validated numerically. As for the third-order isotropic term,
when and only when it is considered, accurate {\it continuum form}
pressure tensor can be definitely obtained, by which the predicted
coexistence densities always agree well with the numerical results.
Compared with this {\it continuum form} pressure tensor, the classical
{\it discrete form} pressure tensor is accurate only when the isotropic
term is a specific one. At last, in the framework of the present
third-order analysis, a consistent scheme for third-order additional term
is proposed, which can be used to independently adjust the coexistence
densities and surface tension. Numerical tests are subsequently carried
out to validate the present scheme.
\end{abstract}


\end{frontmatter}

\section{Introduction} \label{sec-introduction}
Multiphase flows are widely encountered in lots of natural and engineering
systems, such as falling raindrop, cloud formation, droplet-based
microfluidic, phase-change device, etc. Due to the existence of the
deformable phase interface whose position is unknown in advance, numerical
simulation of multiphase flow is much more complicated than that of
single-phase flow. As a powerful and attractive mesoscopic approach for
simulating complex fluid flow problem, the lattice Boltzmann (LB) method has
been applied to the simulation of multiphase flow in past years
\cite{van-der-Graaf2006, Hazi2009, Ledesma-Aguilar2014, Li2016-Review}.
Generally, the existing LB methods for multiphase flow can be grouped into
four major categories: (1) the color-gradient LB method
\cite{Gunstensen1991, Grunau1993, Latva-Kokko2005, Liu2012}, (2) the
pseudopotential LB method \cite{Shan1993, Shan1994, Sbragaglia2007, Li2012,
Khajepor2015}, (3) the free-energy LB method \cite{Swift1995, Swift1996,
Inamuro2000, Pooley2008}, and (4) the kinetic-theory-based LB method
\cite{Luo1998, He2002, McCracken2005, Kikkinides2008}. Among these LB
methods, the pseudopotential LB method, originally proposed by Shan and Chen
\cite{Shan1993, Shan1994}, is the simplest one in both concept and
computation, and thus becomes particularly popular in the LB community for
the simulation of multiphase flow.
\par

In the pseudopotential LB model for multiphase flow, an interaction force is
introduced to mimic the underlying intermolecular interactions, which are
responsible for the formation of multiphase flow. Consequently, phase
transition or separation can be automatically achieved, and thus the
conventional interface capturing and tracking methods are avoided.
Essentially speaking, the interaction force, which is incorporated into the
lattice Boltzmann equation (LBE) through a general forcing scheme, can be
viewed as a finite-difference gradient operator to recover the non-ideal gas
component of the non-monotonic equation of state (EOS) \cite{Shan2006}
(i.e., $p_\text{\tiny EOS}^{} - p^\text{\tiny ideal}$, where $p_\text{\tiny
EOS}^{}$ and $p^\text{\tiny ideal}$ denote the non-monotonic EOS and its
ideal gas component, respectively). Simultaneously, the interfacial
dynamics, such as the non-zero surface tension, are automatically produced
by the higher-order terms in the finite-difference gradient operator. Due to
such simple and integrated treatments of the interfacial dynamics, some
well-known drawbacks exist in the pseudopotential LB model, though its
application has been particularly fruitful \cite{Chibbaro2008, Clime2009,
Yu2009, Varagnolo2013, Li2016, Sun2016}.
\par

One drawback of the pseudopotential LB model is the relatively large
spurious current near the curved phase interface, especially at a large
density ratio. Shan \cite{Shan2006} argued that the spurious current is
caused by the insufficient isotropy of the interaction force (as a
finite-difference gradient operator), and inferred that the spurious current
can be made arbitrarily small by increasing the degree of isotropy of the
interaction force, which is realized by counting the interactions beyond
nearest-neighbor. Numerical tests show the spurious current is suppressed to
some extent by Shan's method \cite{Shan2006, Sbragaglia2007}, and counting
more neighbors will complicate the boundary condition treatment. Sbragaglia
et al. \cite{Sbragaglia2007} investigated the refinement of phase interface
and found that the spurious current can be remarkably reduced by widening
the phase interface (in lattice units). Afterwards, some more methods were
proposed to adjust the interface thickness \cite{Wagner2007, Huang2011,
Li2013-Forcing}. Recently, Guo et al. \cite{Guo2011} and Xiong and Guo
\cite{Xiong2014} analyzed the force balance condition at the discrete
lattice level of LBE, and found that the spurious current is partly caused
by the intrinsic force imbalance in the LBE. Besides the above works, some
other researches have also been made to shed light on the origin of the
spurious current \cite{Wagner2003} and to provide way to reduce the spurious
current \cite{Yu2010}.
\par

Another two drawbacks of the pseudopotential LB model are the thermodynamic
inconsistency (the coexistence densities are inconsistent with the
thermodynamic results) and the nonadjustable surface tension (the surface
tension cannot be adjusted independently of the coexistence densities). Both
of these two drawbacks stem from the simple and integrated treatments of the
interfacial dynamics, since the coexistence densities and surface tension
are affected, or even determined, by the higher-order terms in the
interaction force. In the pseudopotential LB community, it has been widely
shown that different forcing schemes for incorporating the interaction force
into LBE yield distinctly different coexistence densities (particularly the
gas density at a large density ratio) \cite{wagner2006, Huang2011, Sun2012,
Zarghami2015}. Li et al. \cite{Li2012} found that the rationale behind this
phenomenon is that different forcing schemes produce different additional
terms in the recovered macroscopic equation, which have important influences
on the interfacial dynamics for multiphase flow, and then they proposed a
forcing scheme to alleviate the thermodynamic inconsistency. Following the
similar way, some other forcing schemes have been proposed recently
\cite{Li2013-Forcing, Hu2015, Lycett-Brown2015}. As compared to the
thermodynamic inconsistency, the nonadjustable surface tension has not
received much attention. In 2007, Sbragaglia et al. \cite{Sbragaglia2007}
first proposed a multirange pseudopotential LB model, where the surface
tension can be adjusted independently of the EOS. However, as shown by Huang
et al.'s numerical tests \cite{Huang2011}, the coexistence densities, which
are not only determined by the EOS but also affected by the interfacial
dynamics, still vary with the adjustment of surface tension. By introducing
a source term into LBE to incorporate specific additional term, Li and Luo
\cite{Li2013-Tension} proposed a nearest-neighbor-based approach to adjust
the surface tension independently of the coexistence densities. Similar
additional term was also utilized to independently adjust the surface
tension in the latter work by Lycett-Brown and Luo \cite{Lycett-Brown2015}.
\par

Up to date, the above drawbacks in the pseudopotential LB model have been
widely investigated and the corresponding theoretical foundations for the
pseudopotential LB model have been further consolidated. However, there
still exist some theoretical aspects unclear or inconsistent in the
pseudopotential LB model. The isotropic property of the LBE has not been
investigated although this aspect of the interaction force has been clearly
clarified. Accurate pressure tensor cannot be obtained from the recovered
macroscopic equation and the reason is still unclear. Some additional terms,
like $\nabla \cdot (h \mathbf{FF})$ ($h$ is a coefficient and $\mathbf{F}$
is the interaction force), should be recovered at the third-order through
the Chapman-Enskog analysis, but such terms were inconsistently recovered at
the second-order previously. To understand these unclear or inconsistent
theoretical aspects, the traditional second-order Chapman-Enskog analysis,
which is adopted in nearly all previous works, is insufficient, and
higher-order analysis is required. In this work, we target on these
theoretical aspects, and perform a third-order Chapman-Enskog analysis of
the multiple-relaxation-time (MRT) pseudopotential LB model for multiphase
flow. The remainder of the present paper is organized as follows. Section
\ref{sec-lbm} briefly introduces the MRT pseudopotential LB model. Section
\ref{sec-2nd} gives the standard second-order Chapman-Enskog analysis. In
Section \ref{sec-3rd}, a third-order Chapman-Enskog analysis of the MRT
pseudopotential LB model is performed. In Section \ref{sec-results}, the
theoretical results of the third-order analysis are discussed detailedly and
validated numerically. In Section \ref{sec-scheme}, a consistent scheme for
third-order additional term is proposed to independently adjust the
coexistence densities and surface tension. At last, a brief conclusion is
drawn in Section \ref{sec-conclu}.
\par

\section{MRT pseudopotential LB model} \label{sec-lbm}
Without loss of generality, a two-dimensional nine-velocity (D2Q9) MRT
pseudopotential LB model is considered in this work. In the D2Q9 lattice,
discrete velocities are given as
\begin{equation}
\mathbf{e}_i =
\begin{cases}
c \big( 0, \; 0 \big) ^\text{T}, & i=0, \\
c \big( \cos[(i-1)\pi/2], \; \sin[(i-1)\pi/2] \big) ^\text{T},
& i=1,2,3,4, \\
\sqrt{2}c \big( \cos[(2i-1)\pi/4], \; \sin[(2i-1)\pi/4] \big) ^\text{T},
& i=5,6,7,8, \\
\end{cases}
\end{equation}
where $c = \delta_x / \delta_t$ is the lattice speed, and $\delta_x$ and
$\delta_t$ are the lattice spacing and time step, respectively. The MRT LBE
for the density distribution function $\mathbf{f} (\mathbf{x}, t) = \big[
f_0(\mathbf{x}, t), \, \cdots, \, f_8(\mathbf{x}, t) \big] ^\text{T}$ can be
decomposed into two sub-steps: the collision step and the streaming step.
Generally, the collision step is carried out in the moment space
\begin{equation} \label{eq-collision}
\bar{\mathbf{m}}(\mathbf{x}, t) = \mathbf{m}(\mathbf{x}, t) - \mathbf{S}
\left[ \mathbf{m}(\mathbf{x}, t) - \mathbf{m}^\text{eq}(\mathbf{x}, t) \right]
+ \delta_t \left( \mathbf{I} - \dfrac{\mathbf{S}}{2} \right)
\mathbf{F}_m(\mathbf{x}, t),
\end{equation}
while the streaming step is carried out in the velocity space
\begin{equation} \label{eq-streaming}
f_i(\mathbf{x} + \mathbf{e}_i \delta_t, t + \delta_t) =
\bar{f}_i (\mathbf{x}, t).
\end{equation}
Here, $\mathbf{m}(\mathbf{x}, t) = \big[ m_0(\mathbf{x}, t), \, \cdots, \,
m_8(\mathbf{x}, t) \big] ^\text{T} = \mathbf{M} \mathbf{f}(\mathbf{x}, t)$
is the rescaled moment, $\bar{\mathbf{f}} (\mathbf{x}, t) = \big[
\bar{f}_0(\mathbf{x}, t), \, \cdots, \, \bar{f}_8(\mathbf{x}, t) \big]
^\text{T} = \mathbf{M}^{-1} \bar{\mathbf{m}} (\mathbf{x}, t)$ is
post-collision distribution function, $\mathbf{S} = \text{diag} (s_0^{},
s_e, s_\varepsilon, s_j, s_q, s_j, s_q, s_p, s_p)$ is the diagonal
relaxation matrix, $\mathbf{I}$ is the unit matrix, $\mathbf{m}^\text{eq}
(\mathbf{x}, t)$ is the equilibrium moment, and $\mathbf{F}_m (\mathbf{x},
t)$ is the discrete force term. For the D2Q9 lattice, the dimensionless
orthogonal transformation matrix $\mathbf{M}$ can be chosen as
\cite{Lallemand2000}
\begin{equation}
\mathbf{M}=\left( \begin{array}{ccccccccc}
 1& 1& 1& 1& 1& 1& 1& 1& 1\\
-4&-1&-1&-1&-1& 2& 2& 2& 2\\
 4&-2&-2&-2&-2& 1& 1& 1& 1\\
 0& 1& 0&-1& 0& 1&-1&-1& 1\\
 0&-2& 0& 2& 0& 1&-1&-1& 1\\
 0& 0& 1& 0&-1& 1& 1&-1&-1\\
 0& 0&-2& 0& 2& 1& 1&-1&-1\\
 0& 1&-1& 1&-1& 0& 0& 0& 0\\
 0& 0& 0& 0& 0& 1&-1& 1&-1
\end{array} \right).
\end{equation}
Different from previous MRT pseudopotential LB models \cite{Yu2010,
Li2013-Forcing, Hu2015}, and following the pioneering work by Lallemand and
Luo \cite{Lallemand2000}, a free parameter $\alpha$ is retained in the
equilibrium moment $\mathbf{m}^\text{eq} (\mathbf{x}, t)$ as follows
\begin{equation} \label{eq-meq}
\mathbf{m}^\text{eq} = \left( \rho, \;
-2 \rho + 3 \rho \dfrac{|\mathbf{u}|^2}{c^2}, \;
\alpha \rho - 3 \rho \dfrac{|\mathbf{u}|^2}{c^2}, \;
\rho \dfrac{u_x}{c}, \; -\rho \dfrac{u_x}{c}, \;
\rho \dfrac{u_y}{c}, \; -\rho \dfrac{u_y}{c}, \;
\rho \dfrac{u_x^2-u_y^2}{c^2}, \; \rho \dfrac{u_xu_y}{c^2} \right) ^\text{T}.
\end{equation}
Note that the present equilibrium moment degenerates to the classical one
adopted in previous works when $\alpha=1$. The discrete force term in the
moment space $\mathbf{F}_m (\mathbf{x}, t)$ is given as \cite{McCracken2005,
Guo2008}
\begin{equation} \label{eq-fm}
\mathbf{F}_m = \left( 0, \; 6\dfrac{\mathbf{F}\cdot\mathbf{u}}{c^2}, \;
-6\dfrac{\mathbf{F}\cdot\mathbf{u}}{c^2}, \;
\dfrac{F_x}{c}, \; -\dfrac{F_x}{c}, \; \dfrac{F_y}{c}, \; -\dfrac{F_y}{c}, \;
2\dfrac{F_xu_x-F_yu_y}{c^2}, \; \dfrac{F_xu_y+F_yu_x}{c^2} \right) ^\text{T}.
\end{equation}
The macroscopic variables, density $\rho$ and velocity $\mathbf{u}$, are
defined as
\begin{equation} \label{eq-rho-u}
\rho = \sum\limits_{i=0}^{8} f_i, \qquad
\rho \mathbf{u} = \sum\limits_{i=0}^{8} \mathbf{e}_i f_i
+ \dfrac{\delta_t}{2} \mathbf{F}.
\end{equation}
For the above LB model with a force term, it is well known that no
additional term exists in the recovered macroscopic equation at the
Navier-Stokes level \cite{Guo2002}, as will be shown in Section
\ref{sec-2nd}.
\par

In the pseudopotential LB model for multiphase flow, the non-monotonic
equation of state and the non-zero surface tension are simultaneously
produced by the introduction of an interaction force. For the
nearest-neighbor interactions on D2Q9 lattice, the interaction force can be
expressed as \cite{Shan1994, Shan2008}
\begin{equation} \label{eq-interaction-force}
\mathbf{F}(\mathbf{x}) = -G \psi(\mathbf{x}) \sum\limits_{i=1}^{8}
\omega(|\mathbf{e}_i\delta_t|^2) \psi(\mathbf{x} + \mathbf{e}_i\delta_t)
\mathbf{e}_i\delta_t,
\end{equation}
where $\psi(\mathbf{x})$ is the interaction potential (also named as the
pseudopotential), $G$ is the interaction strength, and
$\omega(|\mathbf{e}_i\delta_t|^2)$ are the weights, which are given as
$\omega(\delta_x^2) = 1/3$ and $\omega(2\delta_x^2) = 1/12$ to make
$\mathbf{F}(\mathbf{x})$ fourth-order isotropic \cite{Shan2008}.
Consequently, the following non-monotonic EOS can be obtained
\cite{Shan1994}
\begin{equation} \label{eq-eos}
p_{\text{\tiny EOS}}^{} = \dfrac{\rho c^2}{3}
+ \dfrac{G \delta_x^2}{2} \psi^2,
\end{equation}
where $\rho c^2 /3$ is the ideal gas component ($p^\text{\tiny ideal}$)
recovered by the LBE. For a prescribed EOS in real application, the
interaction potential is inversely calculated by Eq. (\ref{eq-eos}), i.e.,
$\psi = \sqrt{ \raise0.3ex\hbox{$ 2 (p_{\text{\tiny EOS}}^{} - \rho c^2 / 3)
$} \left/ \lower0.3ex\hbox{$ (G \delta_x^2) $} \right. }$. In this case, $G$
can be chosen arbitrarily as long as the term inside the square root is
positive \cite{Yuan2006}. In the present work, the Carnahan-Starling EOS in
thermodynamic theory is taken as an example, which is given as
\cite{Yuan2006, Kupershtokh2009}
\begin{equation}
p_{\text{\tiny EOS}}^{} = K \left[ \rho R T \dfrac{1 + b\rho/4 + (b\rho/4)^2
- (b\rho/4)^3}{ (1-b\rho/4)^3 } - a \rho^2 \right],
\end{equation}
where $R$ is the gas constant, $T$ is the temperature, and $a = 0.4963 R^2
T_c^2 / p_c^{}$ and $b = 0.18727 R T_c / p_c^{}$ with $T_c$ and $p_c^{}$
being the critical temperature and pressure, respectively. Moreover, a
scaling factor $K$ is also included in the EOS, which can be used to adjust
the interface thickness in the simulation \cite{Wagner2007, Hu2015}.
\par

\section{Second-order analysis} \label{sec-2nd}
To establish a starting point for the third-order Chapman-Enskog analysis,
we first perform the standard second-order Chapman-Enskog analysis of the
MRT pseudopotential LB model in this section. Through a second-order Taylor
series expansion of $f_i (\mathbf{x} + \mathbf{e}_i \delta_t, t + \delta_t)$
centered at $(\mathbf{x},t)$, the streaming step (i.e., Eq.
(\ref{eq-streaming})) can be written as
\begin{equation} \label{eq-2nd-taylor-f}
f_i + \delta_t (\partial_t + \mathbf{e}_i \cdot \nabla) f_i +
\dfrac{\delta_t^2}{2} (\partial_t + \mathbf{e}_i \cdot \nabla)^2 f_i +
O(\delta_t^3) = \bar{f}_i.
\end{equation}
Transforming Eq. (\ref{eq-2nd-taylor-f}) into the moment space, and then
combining it with the collision step (i.e., Eq. (\ref{eq-collision})), we
obtain
\begin{equation} \label{eq-2nd-taylor-lbe}
(\mathbf{I} \partial_t + \mathbf{D}) \mathbf{m} +
\dfrac{\delta_t}{2} (\mathbf{I} \partial_t + \mathbf{D})^2 \mathbf{m} +
O(\delta_t^2) =
-\dfrac{\mathbf{S}}{\delta_t} (\mathbf{m} - \mathbf{m}^\text{eq}) +
\left( \mathbf{I} - \dfrac{\mathbf{S}}{2} \right) \mathbf{F}_m,
\end{equation}
where $\mathbf{D} = \mathbf{M} \big[ \text{diag} (\mathbf{e}_0 \cdot \nabla,
\, \cdots, \, \mathbf{e}_8 \cdot \nabla) \big] \mathbf{M}^{-1}$. Eq.
(\ref{eq-2nd-taylor-lbe}) is called the Taylor series expansion of the MRT
LBE in the moment space. Introducing the following Chapman-Enskog expansions
\cite{Guo2007}
\begin{equation} \label{eq-ce-expansions}
\partial_t = \sum\limits_{n=1}^{+\infty} \varepsilon^n \partial_{tn}, \qquad
\nabla = \varepsilon \nabla_1, \qquad
f_i = \sum\limits_{n=0}^{+\infty} \varepsilon^n f_i^{(n)}, \qquad
\mathbf{F} = \varepsilon \mathbf{F}^{(1)},
\end{equation}
there have $\mathbf{D} = \varepsilon \mathbf{D}_1$, $\mathbf{m} =
\sum\nolimits_{n=0}^{+\infty} \varepsilon^n \mathbf{m}^{(n)}$, and
$\mathbf{F}_m = \varepsilon \mathbf{F}_m^{(1)}$, where $\varepsilon$ is the
small expansion parameter. Substituting these Chapman-Enskog expansions into
Eq. (\ref{eq-2nd-taylor-lbe}), we can rewrite Eq. (\ref{eq-2nd-taylor-lbe})
in the consecutive orders of $\varepsilon$ as
\begin{subequations} \label{eq-2nd-orders-lbe}
\begin{equation} \label{eq-2nd-0th-lbe}
\varepsilon^0: \; \mathbf{m}^{(0)} = \mathbf{m}^\text{eq},
\end{equation}
\begin{equation} \label{eq-2nd-1st-lbe}
\varepsilon^1: \; (\mathbf{I} \partial_{t1} + \mathbf{D}_1) \mathbf{m}^{(0)} -
\mathbf{F}_m^{(1)} = - \dfrac{\mathbf{S}}{\delta_t} \left( \mathbf{m}^{(1)} +
\dfrac{\delta_t}{2} \mathbf{F}_m^{(1)} \right),
\end{equation}
\begin{equation} \label{eq-2nd-2nd-lbe}
\varepsilon^2: \; \partial_{t2} \mathbf{m}^{(0)} + (\mathbf{I} \partial_{t1}
+ \mathbf{D}_1) \left( \mathbf{I} - \dfrac{\mathbf{S}}{2} \right)
\left( \mathbf{m}^{(1)} + \dfrac{\delta_t}{2} \mathbf{F}_m^{(1)} \right)
= - \dfrac{\mathbf{S}}{\delta_t} \mathbf{m}^{(2)},
\end{equation}
\end{subequations}
where the first-order ($\varepsilon^1$) equation has been used to simplify
the second-order ($\varepsilon^2$) equation.
\par

To deduce the macroscopic equation, we extract the equations for the
conserved moments ($m_0$, $m_3$, and $m_5$) from Eq.
(\ref{eq-2nd-orders-lbe}) as
\begin{subequations} \label{eq-2nd-orders-m035}
\begin{equation} \label{eq-2nd-0th-m035}
\varepsilon^0: \; \begin{cases}
m_0^{(0)} = m_0^\text{eq}, \\
m_3^{(0)} = m_3^\text{eq}, \\
m_5^{(0)} = m_5^\text{eq},
\end{cases}
\end{equation}
\begin{equation} \label{eq-2nd-1st-m035}
\varepsilon^1: \; \begin{cases}
\partial_{t1} m_0^{(0)} + c \partial_{x1} m_3^{(0)} +
c \partial_{y1} m_5^{(0)} - F_{m0}^{(1)} =  - \tfrac{s_0^{}}{\delta_t}
\left( m_0^{(1)} + \tfrac{\delta_t}{2} F_{m0}^{(1)} \right), \\
\partial_{t1} m_3^{(0)} + c \partial_{x1} \left( \tfrac23 m_0^{(0)} +
\tfrac16 m_1^{(0)} + \tfrac12 m_7^{(0)} \right) + c \partial_{y1} m_8^{(0)}
- F_{m3}^{(1)} = - \tfrac{s_j}{\delta_t} \left( m_3^{(1)}
+ \tfrac{\delta_t}{2} F_{m3}^{(1)} \right), \\
\partial_{t1} m_5^{(0)} + c \partial_{x1} m_8^{(0)} + c \partial_{y1}
\left( \tfrac23 m_0^{(0)} + \tfrac16 m_1^{(0)} - \tfrac12 m_7^{(0)} \right)
- F_{m5}^{(1)} = - \tfrac{s_j}{\delta_t} \left( m_5^{(1)}
+ \tfrac{\delta_t}{2} F_{m5}^{(1)} \right),
\end{cases}
\end{equation}
\begin{equation} \label{eq-2nd-2nd-m035}
\varepsilon^2: \; \begin{cases}
\left( \begin{aligned}
& \partial_{t2} m_0^{(0)} + \partial_{t1} \left( 1-\tfrac{s_0^{}}{2} \right)
\left( m_0^{(1)} + \tfrac{\delta_t}{2} F_{m0}^{(1)} \right) + \\
& c \partial_{x1} \left( 1 - \tfrac{s_j}{2} \right)
\left( m_3^{(1)} + \tfrac{\delta_t}{2} F_{m3}^{(1)} \right)
+ c \partial_{y1} \left( 1 - \tfrac{s_j}{2} \right)
\left( m_5^{(1)} + \tfrac{\delta_t}{2} F_{m5}^{(1)} \right)
\end{aligned} \right)
= - \tfrac{s_0^{}}{\delta_t} m_0^{(2)}, \\
\left( \begin{aligned}
& \partial_{t2} m_3^{(0)} + \partial_{t1} \left( 1 - \tfrac{s_j}{2} \right)
\left( m_3^{(1)} + \tfrac{\delta_t}{2} F_{m3}^{(1)} \right) +
c \partial_{y1} \left( 1 - \tfrac{s_p}{2} \right)
\left( m_8^{(1)} + \tfrac{\delta_t}{2} F_{m8}^{(1)} \right) + \\
& c \partial_{x1} \left[ \tfrac23 \left( 1 - \tfrac{s_0^{}}{2} \right)
\left( m_0^{(1)} + \tfrac{\delta_t}{2} F_{m0}^{(1)} \right) + \tfrac16
\left( 1 - \tfrac{s_e}{2} \right)
\left( m_1^{(1)} + \tfrac{\delta_t}{2} F_{m1}^{(1)} \right) + \tfrac12
\left( 1 - \tfrac{s_p}{2} \right)
\left( m_7^{(1)} + \tfrac{\delta_t}{2} F_{m7}^{(1)} \right)
\right]
\end{aligned} \right)
= - \tfrac{s_j}{\delta_t} m_3^{(2)}, \\
\left( \begin{aligned}
& \partial_{t2} m_5^{(0)} + \partial_{t1} \left( 1 - \tfrac{s_j}{2} \right)
\left( m_5^{(1)} + \tfrac{\delta_t}{2} F_{m5}^{(1)} \right) +
c \partial_{x1} \left( 1 - \tfrac{s_p}{2} \right)
\left( m_8^{(1)} + \tfrac{\delta_t}{2} F_{m8}^{(1)} \right) + \\
& c \partial_{y1} \left[ \tfrac23 \left( 1 - \tfrac{s_0^{}}{2} \right)
\left( m_0^{(1)} + \tfrac{\delta_t}{2} F_{m0}^{(1)} \right) + \tfrac16
\left( 1 - \tfrac{s_e}{2} \right)
\left( m_1^{(1)} + \tfrac{\delta_t}{2} F_{m1}^{(1)} \right) - \tfrac12
\left( 1 - \tfrac{s_p}{2} \right)
\left( m_7^{(1)} + \tfrac{\delta_t}{2} F_{m7}^{(1)} \right)
\right]
\end{aligned} \right)
= - \tfrac{s_j}{\delta_t} m_5^{(2)}.
\end{cases}
\end{equation}
\end{subequations}
Considering $m_0 = \rho$, $m_3 = \rho u_x / c - \tfrac{\delta_t}{2} F_x /c$,
and $m_5 = \rho u_y / c - \tfrac{\delta_t}{2} F_y /c$ (see Eq.
(\ref{eq-rho-u})), Eq. (\ref{eq-2nd-0th-m035}) indicates that
\begin{equation} \label{eq-2nd-0nd-m035-results}
\begin{cases}
m_0^{(1)} + \tfrac{\delta_t}{2} F_{m0}^{(1)}=0, \qquad
m_0^{(n)} = 0 \; (\forall n \geq 2), \\
m_3^{(1)} + \tfrac{\delta_t}{2} F_{m3}^{(1)}=0, \qquad
m_3^{(n)} = 0 \; (\forall n \geq 2), \\
m_5^{(1)} + \tfrac{\delta_t}{2} F_{m5}^{(1)}=0, \qquad
m_5^{(n)} = 0 \; (\forall n \geq 2).
\end{cases}
\end{equation}
Therefore, the first-order equation (i.e., Eq. (\ref{eq-2nd-1st-m035})) can
be simplified as
\begin{equation} \label{eq-2nd-1st-macro}
\varepsilon^1: \; \begin{cases}
\partial_{t1}\rho + \partial_{x1} (\rho u_x) + \partial_{y1} (\rho u_y) =0,\\
\partial_{t1}(\rho u_x) + \partial_{x1} (\rho u_x^2) +
\partial_{y1} (\rho u_x u_y)
= -\partial_{x1} ( \tfrac13 \rho c^2 ) + F_x^{(1)},\\
\partial_{t1}(\rho u_y) + \partial_{x1} (\rho u_x u_y) +
\partial_{y1} (\rho u_y^2)
= -\partial_{y1} ( \tfrac13 \rho c^2 ) + F_y^{(1)}.
\end{cases}
\end{equation}
Based on Eq. (\ref{eq-2nd-1st-macro}), the following relation can be
obtained
\begin{equation} \label{eq-2nd-1st-rhouu}
\begin{split}
\partial_{t1} (\rho \mathbf{uu}) & = [\partial_{t1} (\rho \mathbf{u})]
\mathbf{u} + \mathbf{u} [\partial_{t1} (\rho \mathbf{u})] -
\mathbf{uu} (\partial_{t1} \rho) \\
& = -\tfrac13 c^2 [ (\nabla_1 \rho) \mathbf{u}
+ \mathbf{u} (\nabla_1 \rho) ] + \mathbf{F}^{(1)} \mathbf{u} + \mathbf{u}
\mathbf{F}^{(1)} + O(|\mathbf{u}|^3),
\end{split}
\end{equation}
where the cubic term of velocity will be neglected with the low Mach number
condition. In order to simplify the second-order equation (i.e., Eq.
(\ref{eq-2nd-2nd-m035})), the involved first-order terms on the
non-conserved moments, i.e., $m_1^{(1)} + \tfrac{\delta_t}{2} F_{m1}^{(1)}$,
$m_7^{(1)} + \tfrac{\delta_t}{2} F_{m7}^{(1)}$, and $m_8^{(1)} +
\tfrac{\delta_t}{2} F_{m8}^{(1)}$, should be calculated firstly. These
first-order terms are obtained from Eq. (\ref{eq-2nd-1st-lbe}) and then
simplified with the aid of Eqs. (\ref{eq-2nd-0th-lbe}) and
(\ref{eq-2nd-1st-rhouu}) as:
\begin{subequations} \label{eq-2nd-1st-m178-results}
\begin{equation}
\begin{split}
- \tfrac{s_e}{\delta_t} \left( m_1^{(1)} + \tfrac{\delta_t}{2} F_{m1}^{(1)}
\right) & = \partial_{t1} m_1^{(0)} + c \partial_{x1} \left( m_3^{(0)}
+ m_4^{(0)} \right) + c \partial_{y1} \left( m_5^{(0)} + m_6^{(0)} \right)
- F_{m1}^{(1)} \\
& \approx 2 \rho ( \partial_{x1} u_x + \partial_{y1} u_y ),
\end{split}
\end{equation}
\begin{equation}
\begin{split}
- \tfrac{s_p}{\delta_t} \left( m_7^{(1)} + \tfrac{\delta_t}{2} F_{m7}^{(1)}
\right) &= \partial_{t1} m_7^{(0)} + c \partial_{x1} \left( \tfrac13 m_3^{(0)}
- \tfrac13 m_4^{(0)} \right) - c \partial_{y1} \left( \tfrac13 m_5^{(0)}
- \tfrac13 m_6^{(0)} \right) - F_{m7}^{(1)} \\
& \approx \tfrac23 \rho ( \partial_{x1} u_x - \partial_{y1} u_y ),
\end{split}
\end{equation}
\begin{equation}
\begin{split}
- \tfrac{s_p}{\delta_t} \left( m_8^{(1)} + \tfrac{\delta_t}{2} F_{m8}^{(1)}
\right) &= \partial_{t1} m_8^{(0)} + c \partial_{x1} \left( \tfrac23 m_5^{(0)}
+ \tfrac13 m_6^{(0)} \right) + c \partial_{y1} \left( \tfrac23 m_3^{(0)}
+ \tfrac13 m_4^{(0)} \right) - F_{m8}^{(1)} \\
& \approx \tfrac13 \rho ( \partial_{x1} u_y + \partial_{y1} u_x ),
\end{split}
\end{equation}
\end{subequations}
where the sign `` $\approx$ '' means the cubic term of velocity is
neglected. With the aid of Eqs. (\ref{eq-2nd-0th-lbe}),
(\ref{eq-2nd-0nd-m035-results}), and (\ref{eq-2nd-1st-m178-results}), the
second-order equation (i.e., Eq. (\ref{eq-2nd-2nd-m035})) can be finally
simplified as
\begin{equation} \label{eq-2nd-2nd-macro}
\varepsilon^2: \; \begin{cases}
\partial_{t2} \rho =0, \\
\partial_{t2} (\rho u_x) = \partial_{x1} [\rho \nu (\partial_{x1} u_x -
\partial_{y1} u_y )] + \partial_{y1} [\rho \nu (\partial_{y1} u_x +
\partial_{x1} u_y )] + \partial_{x1} [\rho \varsigma (\partial_{x1} u_x +
\partial_{y1} u_y )], \\
\partial_{t2} (\rho u_y) = \partial_{x1} [\rho \nu (\partial_{x1} u_y +
\partial_{y1} u_x )] + \partial_{y1} [\rho \nu (\partial_{y1} u_y -
\partial_{x1} u_x )] + \partial_{y1} [\rho \varsigma (\partial_{x1} u_x +
\partial_{y1} u_y )],
\end{cases}
\end{equation}
where $\nu = c^2 \delta_t (s_p^{-1} - 0.5) /3$ is the kinetic viscosity,
$\varsigma = c^2 \delta_t (s_e^{-1} - 0.5) /3$ is the bulk viscosity.
Combining the first- and second-order equations (i.e., Eqs.
(\ref{eq-2nd-1st-macro}) and (\ref{eq-2nd-2nd-macro})), the following
macroscopic equation at the Navier-Stokes level (second-order) can be
recovered
\begin{equation} \label{eq-2nd-orders-macro}
\begin{cases}
\partial_t \rho + \nabla \cdot (\rho \mathbf{u}) = 0, \\
\partial_t (\rho \mathbf{u}) + \nabla \cdot (\rho \mathbf{uu}) =
-\nabla(\tfrac13 \rho c^2) + \mathbf{F} + \nabla \cdot \big\{ \rho \nu
[\nabla \mathbf{u} + \mathbf{u} \nabla - (\nabla\cdot\mathbf{u})\mathbf{I} ]
\big\} + \nabla (\rho \varsigma \nabla \cdot \mathbf{u} ).
\end{cases}
\end{equation}
From the above second-order Chapman-Enskog analysis, we can see that the
free parameter $\alpha$ makes no difference to the recovered macroscopic
equation at the Navier-Stokes level. Moreover, the force term is correctly
recovered, i.e., no discrete lattice effect exists.
\par

\section{Third-order analysis} \label{sec-3rd}
To identify the higher-order terms in the recovered macroscopic equation, a
third-order Chapman-Enskog analysis of the MRT pseudopotential LB model is
carried out in this section. Performing the Taylor series expansion of the
streaming step (i.e., Eq. (\ref{eq-streaming})) to third-order, and then
transforming the result into the moment space and combining it with the
collision step (i.e., Eq. (\ref{eq-collision})), the following Taylor series
expansion of the MRT LBE in the moment space can be obtained
\begin{equation} \label{eq-3rd-taylor-lbe}
(\mathbf{I} \partial_t + \mathbf{D}) \mathbf{m} + \dfrac{\delta_t}{2}
(\mathbf{I} \partial_t + \mathbf{D})^2 \mathbf{m} + \dfrac{\delta_t^2}{6}
(\mathbf{I} \partial_t + \mathbf{D})^3 \mathbf{m} + O(\delta_t^3) =
-\dfrac{\mathbf{S}}{\delta_t} (\mathbf{m} - \mathbf{m}^\text{eq}) +
\left( \mathbf{I} - \dfrac{\mathbf{S}}{2} \right) \mathbf{F}_m .
\end{equation}
With the Chapman-Enskog expansions given by Eq. (\ref{eq-ce-expansions}),
Eq. (\ref{eq-3rd-taylor-lbe}) can be rewritten in the consecutive orders of
$\varepsilon$ as
\begin{subequations} \label{eq-3rd-orders-lbe}
\begin{equation} \label{eq-3rd-0th-lbe}
\varepsilon^0: \; \mathbf{m}^{(0)} = \mathbf{m}^\text{eq},
\end{equation}
\begin{equation} \label{eq-3rd-1st-lbe}
\varepsilon^1: \; (\mathbf{I} \partial_{t1} + \mathbf{D}_1) \mathbf{m}^{(0)}-
\mathbf{F}_m^{(1)} = - \dfrac{\mathbf{S}}{\delta_t} \left( \mathbf{m}^{(1)}
+ \dfrac{\delta_t}{2} \mathbf{F}_m^{(1)} \right) ,
\end{equation}
\begin{equation} \label{eq-3rd-2nd-lbe}
\varepsilon^2: \; \partial_{t2} \mathbf{m}^{(0)} + (\mathbf{I} \partial_{t1}
+ \mathbf{D}_1) \mathbf{m}^{(1)} + \dfrac{\delta_t}{2} (\mathbf{I}
\partial_{t1} + \mathbf{D}_1)^2 \mathbf{m}^{(0)} =
-\dfrac{\mathbf{S}}{\delta_t} \mathbf{m}^{(2)},
\end{equation}
\begin{equation} \label{eq-3rd-3rd-lbe}
\varepsilon^3: \; \left( \begin{aligned}
& \partial_{t3} \mathbf{m}^{(0)} + \partial_{t2} \mathbf{m}^{(1)} +
(\mathbf{I} \partial_{t1} + \mathbf{D}_1) \mathbf{m}^{(2)} +
\delta_t (\mathbf{I} \partial_{t1} + \mathbf{D}_1)
\partial_{t2} \mathbf{m}^{(0)} + \\
& \dfrac{\delta_t}{2} (\mathbf{I} \partial_{t1} + \mathbf{D}_1)^2
\mathbf{m}^{(1)} + \dfrac{\delta_t^2}{6}
(\mathbf{I} \partial_{t1} + \mathbf{D}_1)^3 \mathbf{m}^{(0)}
\end{aligned}\right)
= - \dfrac{\mathbf{S}}{\delta_t} \mathbf{m}^{(3)} .
\end{equation}
\end{subequations}
Here, the equations at the orders of $\varepsilon^0$, $\varepsilon^1$, and
$\varepsilon^2$ (i.e., Eqs. (\ref{eq-3rd-0th-lbe}), (\ref{eq-3rd-1st-lbe}),
and (\ref{eq-3rd-2nd-lbe})) are identical to those in the second-order
analysis (i.e., Eqs. (\ref{eq-2nd-0th-lbe}), (\ref{eq-2nd-1st-lbe}), and
(\ref{eq-2nd-2nd-lbe})). Therefore, at the Navier-Stokes level, Eq.
(\ref{eq-2nd-orders-macro}) can also be recovered from Eq.
(\ref{eq-3rd-orders-lbe}). From Eq. (\ref{eq-3rd-orders-lbe}), we can see
that the equation at the order of $\varepsilon^3$ (i.e., Eq.
(\ref{eq-3rd-3rd-lbe})) is much more complicated than the equations at the
lower-order. Proceeding along the general way, deducing the corresponding
macroscopic equation from Eq. (\ref{eq-3rd-3rd-lbe}) is difficult and rather
cumbersome, and will lead to the Burnett level equation. This is clearly
unnecessary and not the desired result in this work.
\par

As it is well known, the second-order Chapman-Enskog analysis is sufficient
for single-phase flow, and the main difference between the single-phase and
multiphase flows is the large density gradient near the phase interface. In
the pseudopotential LB model for multiphase flow, such density gradient is
directly caused by the interaction force and is irrelevant to time and
velocity. Therefore, the goal of the present third-order analysis is to
identify the time- and velocity-independent leading terms on the interaction
force at the third-order. Keeping this goal in mind, we can consider a
steady and stationary situation for the sake of simplicity. For the steady
situation, all the time derivative terms are zero, and then Eq.
(\ref{eq-3rd-orders-lbe}) can be simplified as
\begin{subequations} \label{eq-3rd-orders-lbe-simple}
\begin{equation} \label{eq-3rd-0th-lbe-simple}
\varepsilon^0: \; \mathbf{m}^{(0)} = \mathbf{m}^\text{eq},
\end{equation}
\begin{equation} \label{eq-3rd-1st-lbe-simple}
\varepsilon^1: \; \partial_{t1} \mathbf{m}^{(0)} +
\mathbf{D}_1 \mathbf{m}^{(0)} - \mathbf{F}_m^{(1)} =
- \dfrac{\mathbf{S}}{\delta_t} \left( \mathbf{m}^{(1)} + \dfrac{\delta_t}{2}
\mathbf{F}_m^{(1)} \right),
\end{equation}
\begin{equation} \label{eq-3rd-2nd-lbe-simple}
\varepsilon^2: \; \partial_{t2} \mathbf{m}^{(0)} - \delta_t \mathbf{D}_1
\left( \mathbf{S}^{-1} - \dfrac{\mathbf{I}}{2} \right)
\left( \mathbf{D}_1 \mathbf{m}^{(0)} - \mathbf{F}_m^{(1)} \right) =
- \dfrac{\mathbf{S}}{\delta_t} \mathbf{m}^{(2)},
\end{equation}
\begin{equation} \label{eq-3rd-3rd-lbe-simple}
\varepsilon^3: \; \partial_{t3} \mathbf{m}^{(0)} + \delta_t^2 \left[
\mathbf{D}_1 \left( \mathbf{S}^{-1} - \dfrac{\mathbf{I}}{2} \right)
\mathbf{D}_1 \left( \mathbf{S}^{-1} - \dfrac{\mathbf{I}}{2} \right)
\left( \mathbf{D}_1 \mathbf{m}^{(0)} - \mathbf{F}_m^{(1)} \right) -
\dfrac{1}{12} \mathbf{D}_1^3 \mathbf{m}^{(0)} \right] =
- \dfrac{\mathbf{S}}{\delta_t} \mathbf{m}^{(3)},
\end{equation}
\end{subequations}
where the lower-order equations have been used to simplify the higher-order
equations. Note that the terms $\partial_{t1} \mathbf{m}^{(0)}$,
$\partial_{t2} \mathbf{m}^{(0)}$, and $\partial_{t3} \mathbf{m}^{(0)}$ are
reserved in Eq. (\ref{eq-3rd-orders-lbe-simple}) though they are equal to
zero. These time derivative terms act as a gauge to avoid the wrong scaling
among the equations at different orders. As for the stationary situation,
the velocity is zero, i.e., $\mathbf{u} = \mathbf{0}$.
\par

Similar to the second-order analysis, the equations for the conserved
moments ($m_0$, $m_3$, and $m_5$) are extracted from Eq.
(\ref{eq-3rd-orders-lbe-simple}) to deduce the macroscopic equation. The
zeroth-order ($\varepsilon^0$) equations for the conserved moments in Eq.
(\ref{eq-3rd-0th-lbe-simple}) are
\begin{equation}
\varepsilon^0: \; \begin{cases}
m_0^{(0)} = m_0^\text{eq}, \\
m_3^{(0)} = m_3^\text{eq}, \\
m_5^{(0)} = m_5^\text{eq},
\end{cases}
\end{equation}
which indicates that
\begin{equation} \label{eq-3rd-0nd-m035-results}
\begin{cases}
m_0^{(1)} + \tfrac{\delta_t}{2} F_{m0}^{(1)}=0, \qquad
m_0^{(n)} = 0 \; (\forall n \geq 2), \\
m_3^{(1)} + \tfrac{\delta_t}{2} F_{m3}^{(1)}=0, \qquad
m_3^{(n)} = 0 \; (\forall n \geq 2), \\
m_5^{(1)} + \tfrac{\delta_t}{2} F_{m5}^{(1)}=0, \qquad m_5^{(n)} = 0 \;
(\forall n \geq 2).
\end{cases}
\end{equation}
With the aid of Eqs. (\ref{eq-3rd-0th-lbe-simple}) and
(\ref{eq-3rd-0nd-m035-results}), the first-order ($\varepsilon^1$) equations
for the conserved moments in Eq. (\ref{eq-3rd-1st-lbe-simple}) are
\begin{equation} \label{eq-3rd-1st-macro}
\varepsilon^1: \; \begin{cases}
\partial_{t1} \rho =0 , \\
\partial_{t1} (\rho u_x) = -\partial_{x1} (\tfrac13 \rho c^2)+F_{x}^{(1)}, \\
\partial_{t1} (\rho u_y) = -\partial_{y1} (\tfrac13 \rho c^2)+F_{y}^{(1)}.
\end{cases}
\end{equation}
Similarly, the second-order ($\varepsilon^2$) equations for the conserved
moments in Eq. (\ref{eq-3rd-2nd-lbe-simple}) are
\begin{equation} \label{eq-3rd-2nd-macro}
\varepsilon^2: \; \begin{cases}
\partial_{t2} \rho = 0, \\
\partial_{t2} (\rho u_x) =0, \\
\partial_{t2} (\rho u_y) =0.
\end{cases}
\end{equation}
To simplify the descriptions in the following, we introduce $\text{diag}
(\sigma_0^{}, \sigma_e, \sigma_\varepsilon, \sigma_j, \sigma_q, \sigma_j,
\sigma_q, \sigma_p, \sigma_p) = \mathbf{S}^{-1} - \mathbf{I}/2$. After some
lengthy algebra, the third-order ($\varepsilon^3$) equations for the
conserved moments in Eq. (\ref{eq-3rd-3rd-lbe-simple}) are
\begin{equation} \label{eq-3rd-3rd-macro}
\varepsilon^3: \; \begin{cases}
\partial_{t3} \rho =0, \\
\partial_{t3} (\rho u_x) = -\delta_t^2 c^2 \left[
\dfrac{2 (\alpha - 1) (\sigma_e \sigma_q - \sigma_p \sigma_q) - 1}{12}
\left( \partial_{x1}^2 F_x^{(1)} + \partial_{y1}^2 F_x^{(1)} \right) +
\dfrac{(\alpha - 1)(12\sigma_p \sigma_q - 1)}{12}
\partial_{y1}^2 F_x^{(1)} \right], \\
\partial_{t3} (\rho u_y) = -\delta_t^2 c^2 \left[
\dfrac{2 (\alpha - 1) (\sigma_e \sigma_q - \sigma_p \sigma_q) - 1}{12}
\left( \partial_{x1}^2 F_y^{(1)} + \partial_{y1}^2 F_y^{(1)} \right) +
\dfrac{(\alpha - 1)(12\sigma_p \sigma_q - 1)}{12}
\partial_{x1}^2 F_y^{(1)} \right].
\end{cases}
\end{equation}
Combining the first-, second-, and third-order equations (i.e., Eqs.
(\ref{eq-3rd-1st-macro}), (\ref{eq-3rd-2nd-macro}), and
(\ref{eq-3rd-3rd-macro})) together, we finally obtain the following
third-order macroscopic equation
\begin{equation} \label{eq-3rd-orders-macro}
\begin{cases}
\partial_t \rho =0, \\
\partial_t (\rho \mathbf{u}) = -\nabla(\tfrac13 \rho c^2) + \mathbf{F}
+ \mathbf{R}_\text{iso} + \mathbf{R}_\text{aniso},
\end{cases}
\end{equation}
where $\mathbf{R}_\text{iso}$ and $\mathbf{R}_\text{aniso}$ are the
third-order isotropic and anisotropic terms that are expressed as
\begin{subequations}
\begin{equation} \label{eq-3rd-iso-term}
\mathbf{R}_\text{iso} = -\delta_t^2 c^2 \dfrac{2 (\alpha-1)
(\sigma_e\sigma_q - \sigma_p \sigma_q) -1}{12} \nabla \cdot \nabla \mathbf{F},
\end{equation}
\begin{equation} \label{eq-3rd-aniso-term}
\mathbf{R}_\text{aniso} = -\delta_t^2 c^2 \dfrac{ (\alpha-1)
(12 \sigma_p\sigma_q - 1) }{12} \left( \partial_y^2 F_x, \;
\partial_x^2 F_y \right)^\text{T}.
\end{equation}
\end{subequations}
From the above third-order Chapman-Enskog analysis, we can see that the
time- and velocity-independent leading terms on the interaction force
definitely exist in the recovered macroscopic equation at the third-order,
and the free parameter $\alpha$ has crucial influence on these third-order
terms. Note that the above third-order terms still exist for the general
situation, even though they are identified under a specific condition.
\par

\section{Discussions and validations} \label{sec-results}
In this section, the theoretical results of the present third-order
Chapman-Enskog analysis will be discussed detailedly and validated
numerically. Firstly, the isotropic property of the LBE is investigated,
with a focus on the third-order anisotropic term. Then, the determination of
the pressure tensor, which is of crucial importance for multiphase flow, is
analyzed, with a focus on the third-order isotropic term. For the numerical
validations, the basic simulation parameters are set as $\delta_x = 1$,
$\delta_t = 1$, $G = -1$, $a = 1$, $b = 4$, $R = 1$, and $K = 1$, while the
rest simulation parameters will be given individually for different cases.
\par

\subsection{Isotropic property of the LBE} \label{subsec-isotropy}
\begin{figure}[htbp]
  \centering
  \includegraphics[scale=1,draft=\figdraft]{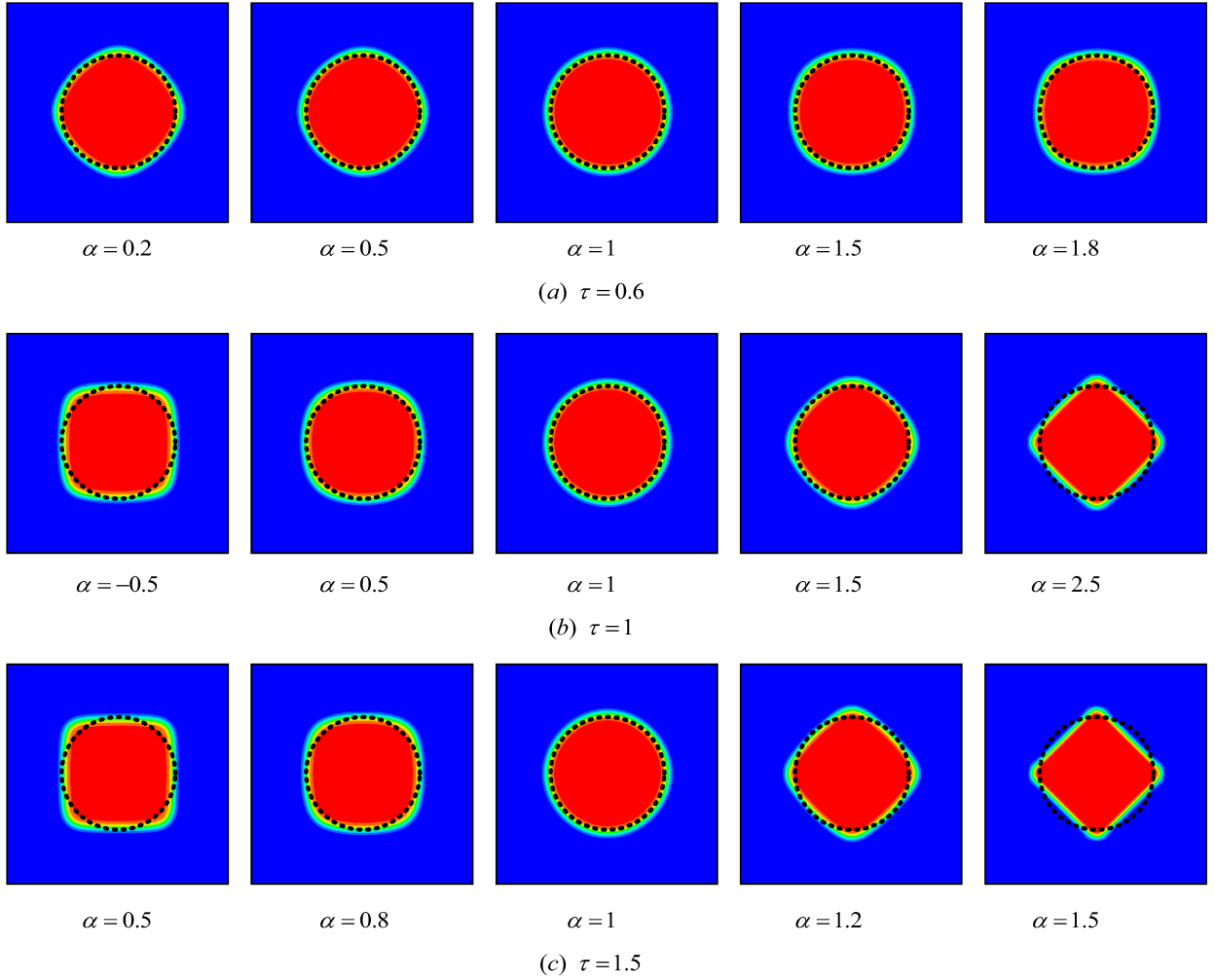}
  \caption{Steady-state density contours of the stationary droplet for
  varied $\alpha$ and different $\tau$. The inserted dashed circle is
  the initial shape of the stationary droplet.}
  \label{fig-droplet-srt}
\end{figure}

At the second-order (Navier-Stokes level), the recovered macroscopic
equation is always isotropic (see Eq. (\ref{eq-2nd-orders-macro})). However,
at the third-order, anisotropic term $\mathbf{R}_\text{aniso}$ is recovered
by the LBE in the macroscopic momentum equation (see Eq.
(\ref{eq-3rd-orders-macro})). To show the effect of such anisotropic term on
multiphase flow, numerical simulations of stationary droplet are carried out
on a $N_x \times N_y = 128 \times 128$ lattice with periodic boundary
conditions in both $x$ and $y$ directions. The relaxation parameters are set
as $s_0^{} = s_j = s_p = s_q = s_e = s_\varepsilon = 1/\tau$. Here, $\tau$
is the dimensionless relaxation time. The temperature is chosen as $T = 0.9
T_c$, which indicates that the thermodynamic gas and liquid densities given
by the Maxwell construction are $\rho_g^\text{thermo} = 4.5435 \times
10^{-2}$ and $\rho_l^\text{thermo} = 2.4806 \times 10^{-1}$, respectively.
In the simulation, the density and velocity fields are initialized as
\begin{subequations} \label{eq-initialize-droplet}
\begin{equation}
\rho (\mathbf{x}) = \dfrac{\rho_g^\text{thermo} + \rho_l^\text{thermo}}{2} +
\dfrac{\rho_g^\text{thermo} - \rho_l^\text{thermo}}{2}
\tanh \dfrac{2(|\mathbf{x}-\mathbf{x}_c|-r_0)}{W},
\end{equation}
\begin{equation}
\mathbf{u} (\mathbf{x}) = \mathbf{0},
\end{equation}
\end{subequations}
where $\mathbf{x}_c = \tfrac{\delta_x}{2} (N_x, \, N_y)^\text{T}$ is the
central position of the computational domain, $W = 5 \delta_x$ is the
initial interface width, and $r_0 = \tfrac{\delta_x}{4} N_x$ is the initial
droplet radius. Fig. \ref{fig-droplet-srt} shows the steady-state density
contours of the droplet for varied $\alpha$ and different $\tau$. It can be
clearly seen that when $\alpha \neq 1$ (i.e., $\mathbf{R}_\text{aniso} \neq
\mathbf{0}$, see Eq. (\ref{eq-3rd-aniso-term})), the droplet becomes
out-of-round and its shape is $\tau\text{-dependent}$; when $\alpha = 1$
(i.e., $\mathbf{R}_\text{aniso} = \mathbf{0}$, see Eq.
(\ref{eq-3rd-aniso-term})), the shape of droplet is independent of $\tau$
and keeps circular consistently. These results suggest that the third-order
anisotropic term recovered by the LBE has important influence on multiphase
flow and must be eliminated in real application, which also indicate that
the isotropy of the LBE should be third-order at least in the
pseudopotential LB model for multiphase flow.
\par

\begin{figure}[htbp]
  \centering
  \includegraphics[scale=1,draft=\figdraft]{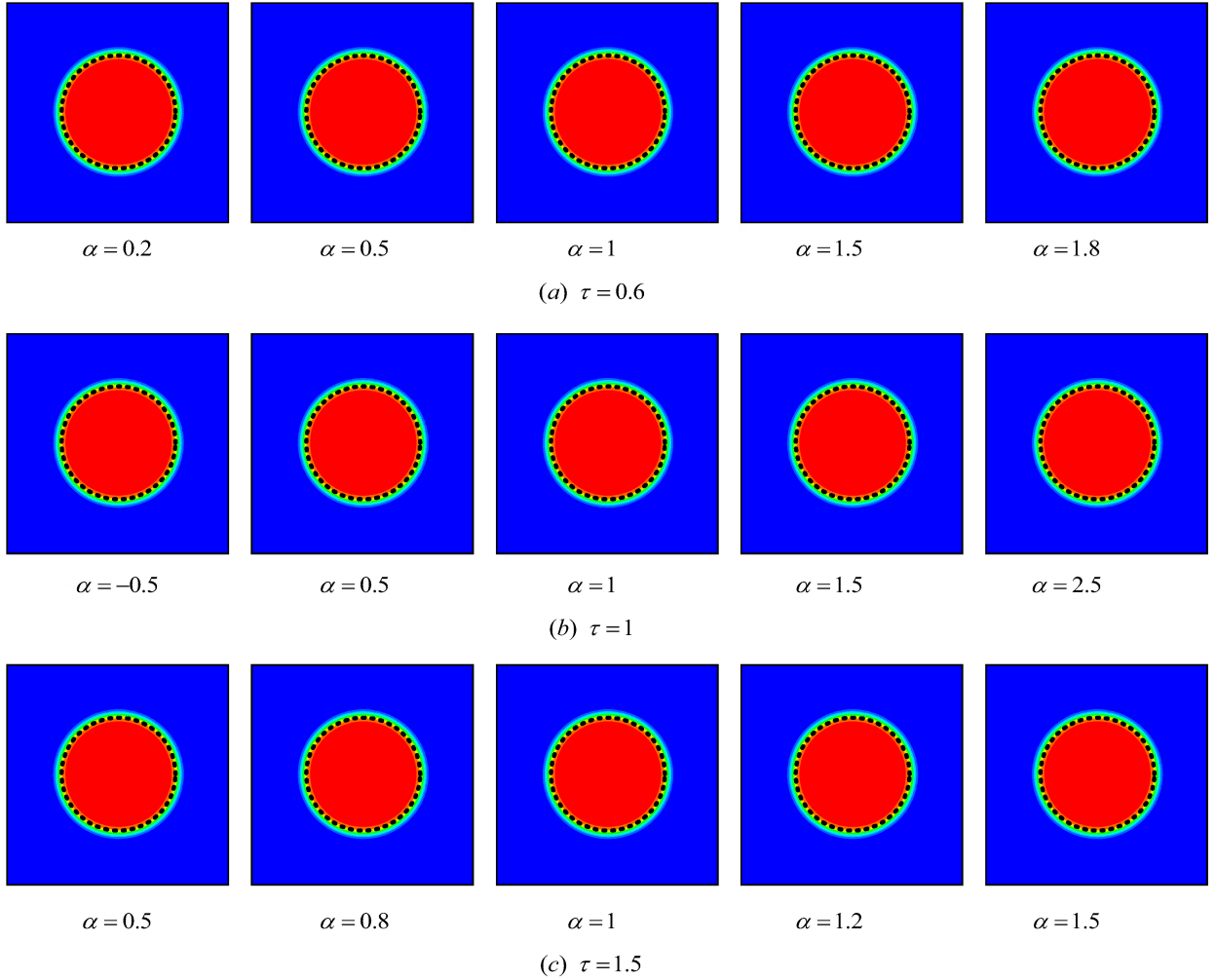}
  \caption{Steady-state density contours of the stationary droplet for
  varied $\alpha$ and different $\tau$ when the ``magic'' parameter
  ${\it \Lambda} \equiv 1/12$. The inserted dashed circle is the initial
  shape of the stationary droplet.}
  \label{fig-droplet-mrt}
\end{figure}

The present third-order analysis shows that the third-order anisotropic term
$\mathbf{R}_\text{aniso}$ is eliminated when $\alpha = 1$ (see Eq.
(\ref{eq-3rd-aniso-term})), which means that the general equilibrium moment
given by Eq. (\ref{eq-meq}) degenerates to the classical one adopted in
previous works. At the same time, it is interesting to find from Eq.
(\ref{eq-3rd-aniso-term}) that by setting a ``magic'' parameter to $1 / 12$
as follows
\begin{equation}
{\it \Lambda} = \sigma_p \sigma_q = \left( \dfrac{1}{s_p} - \dfrac12 \right)
\left( \dfrac{1}{s_q} - \dfrac12 \right) \equiv \dfrac{1}{12},
\end{equation}
the anisotropic term $\mathbf{R}_\text{aniso}$ can also be eliminated. To
validate this point, the same numerical simulations of stationary droplet
are carried out except that the relaxation parameters are set as: $s_0^{} =
s_j = s_p = s_e = s_\varepsilon = 1/ \tau$ and $s_q = 1/ [0.5 + {\it
\Lambda} / (s_p^{-1} - 0.5)]$ with ${\it \Lambda} \equiv 1/12$. The
steady-state density contours of the droplet are shown in Fig.
\ref{fig-droplet-mrt}. As expected, the final shape of droplet is circular
perfectly for all involved $\alpha$ (including $\alpha \neq 1$) and
different $\tau$, which validates the successful elimination of
$\mathbf{R}_\text{aniso}$ by setting ${\it \Lambda} \equiv 1/12$ and also
demonstrates the effectiveness of the present third-order analysis. The
above numerical simulations clearly show the necessity of eliminating the
third-order anisotropic term $\mathbf{R}_\text{aniso}$. Similarly, the
third-order isotropic term $\mathbf{R}_\text{iso}$ needs to be considered as
well, which will be discussed in the next section.
\par

At the end of this section, a further discussion on the isotropic property
of the pseudopotential LB model is deserved. It is well known that the
interaction force given by Eq. (\ref{eq-interaction-force}) is fourth-order
isotropic, and increasing the degree of isotropy of the interaction force
can help to reduce the spurious current \cite{Shan2006}. According to the
discussion on the third-order anisotropic term recovered by the LBE in this
section, the isotropic property of the LBE also has significant influence on
multiphase flow in the pseudopotential LB model. Considering the LBE on D2Q9
lattice can achieve fourth-order isotropy at most, some anisotropic terms
will emerge in the recovered macroscopic equation at the fifth-order, even
though the interaction force is infinite-order isotropic. These higher-order
anisotropic terms intrinsically recovered by the LBE will produce some
spurious current inevitably. Therefore, the spurious current is partly
caused by the finite-order isotropy of the LBE on a discrete lattice, which
was not realized previously, and it cannot be made arbitrarily small just by
increasing the degree of isotropy of the interaction force.
\par

\subsection{Determination of the pressure tensor} \label{subsec-pressure}
In the pseudopotential LB model for multiphase flow, determination of the
pressure tensor is of crucial importance. Many macroscopic properties, such
as the coexistence densities, can be predicted analytically by the pressure
tensor. Generally, the pressure tensor can be determined in two forms: the
{\it continuum form} pressure tensor and the {\it discrete form} pressure
tensor. In the pseudopotential LB community, it is well known that the {\it
continuum form} pressure tensor, which is obtained from the macroscopic
equation recovered through the Chapman-Enskog analysis, is inaccurate in
predicting the macroscopic properties, and the {\it discrete form} pressure
tensor, which is exactly constructed on the discrete lattice, should be used
for the predictions. For the nearest-neighbor interactions given by Eq.
(\ref{eq-interaction-force}), the {\it discrete form} pressure tensor is
given as \cite{Shan2008, Sbragaglia2013}
\begin{equation} \label{eq-p-discrete}
\mathbf{P}^\text{discrete} (\mathbf{x}) = \dfrac{\rho(\mathbf{x}) c^2}{3}
\mathbf{I} + \dfrac{G}{2} \psi (\mathbf{x}) \sum\limits_{i=1}^8
\omega(|\mathbf{e}_i \delta_t|^2) \psi (\mathbf{x} + \mathbf{e}_i \delta_t)
\mathbf{e}_i \delta_t \mathbf{e}_i \delta_t.
\end{equation}
Performing the Taylor series expansion of $\psi (\mathbf{x} + \mathbf{e}_i
\delta_t)$ centered at $\mathbf{x}$, Eq. (\ref{eq-p-discrete}) can be
further expressed as
\begin{equation} \label{eq-p-discrete-limit}
\mathbf{P}^\text{discrete} = \left( \dfrac{\rho c^2}{3} +
\dfrac{G \delta_x^2}{2} \psi^2 + \dfrac{G \delta_x^4}{12}
\psi \nabla \cdot \nabla \psi \right) \mathbf{I} +
\dfrac{G \delta_x^4}{6} \psi \nabla \nabla \psi + O(\nabla^4),
\end{equation}
where the higher-order terms are anisotropic and will be neglected. To
determine the coexistence densities, a steady-state one-dimensional flat
interface along $y$ direction can be considered. Then, the normal pressure
$P_n^\text{discrete}$ given by Eq. (\ref{eq-p-discrete-limit}) is
\begin{equation} \label{eq-pn-discrete-limit}
P_n^\text{discrete} = P_{xx} = \dfrac{\rho c^2}{3} + \dfrac{G \delta_x^2}{2}
\psi^2 + \dfrac{G \delta_x^4}{4} \psi \dfrac{d^2\psi}{dx^2}.
\end{equation}
According to Eq. (\ref{eq-pn-discrete-limit}) and after some algebra, the
following integral equation, which is called the mechanical stability
condition, can be obtained \cite{Shan1994, Shan2008}
\begin{equation} \label{eq-mechanical-discrete}
\int_{\rho_g^{}}^{\rho_l^{}} \left( p_0^{} - \dfrac{\rho c^2}{3} -
\dfrac{G \delta_x^2}{2} \psi^2 \right) \dfrac{\psi'}{\psi^{1+\epsilon}}
d\rho =0 \enskip \text{with} \enskip \epsilon =0,
\end{equation}
where $\psi' = d\psi / d\rho$, and $p_0^{} = p_{\text{\tiny EOS}}^{}
(\rho_g^{}) = p_{\text{\tiny EOS}}^{} (\rho_l^{})$ is the bulk pressure.
Based on Eq. (\ref{eq-mechanical-discrete}), the coexistence densities
($\rho_g^{}$ and $\rho_l^{}$) can be determined analytically via numerical
integration.
\par

With the consideration of the present third-order analysis performed in
Section \ref{sec-3rd}, the {\it continuum form} pressure tensor is defined
as
\begin{equation} \label{eq-p-definition}
\nabla \cdot \mathbf{P} = \nabla ( \tfrac13 \rho c^2 ) - \mathbf{F}
- \mathbf{R}_\text{iso}.
\end{equation}
Compared with previous works \cite{He2002, Sbragaglia2007}, the third-order
isotropic term $\mathbf{R}_\text{iso}$ is considered in the definition. Note
that the third-order anisotropic term $\mathbf{R}_\text{aniso}$ should be
zero as discussed in Section \ref{subsec-isotropy}. Performing the Taylor
series expansion of $\psi (\mathbf{x} + \mathbf{e}_i \delta_t)$ centered at
$\mathbf{x}$, the interaction force $\mathbf{F}$ given by Eq.
(\ref{eq-interaction-force}) can be expressed as
\begin{equation} \label{eq-taylor-force}
\begin{split}
\mathbf{F} & = -G \delta_x^2 \psi \nabla \psi - \dfrac{G \delta_x^4}{6}
\psi \nabla \nabla \cdot \nabla \psi + O(\nabla^5) \\
& = - \dfrac{G \delta_x^2}{2} \nabla \cdot (\psi^2 \mathbf{I}) -
\dfrac{G \delta_x^4}{6} \nabla \cdot \left[ a_1 \nabla \psi \nabla \psi
+ a_2 \psi \nabla \nabla \psi + (a_3 \nabla \psi \cdot \nabla \psi +
a_4 \psi \nabla \cdot \nabla \psi) \mathbf{I} +O(\nabla^4) \right],
\end{split}
\end{equation}
where the higher-order terms are anisotropic and will be neglected, and
$a_{1-4}$ are free parameters that satisfy \cite{Sbragaglia2007}
\begin{equation} \label{eq-parameters-a03}
\begin{cases}
a_1 + a_2 + 2a_3 = 0, \\
a_1 + a_4 = 0, \\
a_2 + a_4 = 1.
\end{cases}
\end{equation}
With the aid of Eq. (\ref{eq-taylor-force}), the third-order isotropic term
$\mathbf{R}_\text{iso}$ given by Eq. (\ref{eq-3rd-iso-term}) can be
expressed as
\begin{equation} \label{eq-taylor-3rd-iso-term}
\begin{split}
\mathbf{R}_\text{iso} & = \delta_t^2 c^2 \dfrac{2 (\alpha-1)
(\sigma_e \sigma_q - \sigma_p \sigma_q) - 1}{12} \dfrac{G \delta_x^2}{2}
\nabla \nabla \cdot \nabla \psi^2 + O(\nabla^5) \\
& = \dfrac{k_d G \delta_x^4}{2} \nabla \cdot \left[ b_1 \nabla \nabla \psi^2
+ b_2 ( \nabla \cdot \nabla \psi^2 ) \mathbf{I} + O(\nabla^4) \right],
\end{split}
\end{equation}
where $k_d = [2 (\alpha-1) (\sigma_e \sigma_q - \sigma_p \sigma_q) - 1] /
12$, and $b_{1-2}$ are free parameters that satisfy
\begin{equation} \label{eq-parameters-b01}
b_1 + b_2 =1.
\end{equation}
Substituting Eqs. (\ref{eq-taylor-force}) and (\ref{eq-taylor-3rd-iso-term})
into Eq. (\ref{eq-p-definition}), and considering $\nabla \nabla \psi^2 = 2
\nabla \psi \nabla \psi + 2 \psi \nabla \nabla \psi$ and $\nabla \cdot
\nabla \psi^2 = 2 \nabla \psi \cdot \nabla \psi + 2 \psi \nabla \cdot \nabla
\psi$, the {\it continuum form} pressure tensor can be finally obtained as
\begin{equation} \label{eq-p}
\begin{split}
\mathbf{P} = \Bigg( \dfrac{\rho c^2}{3} + \dfrac{G \delta_x^2}{2} \psi^2
& + \dfrac{G \delta_x^4}{6} \left[ (a_3 - 6k_d b_2) \nabla \psi \cdot \nabla \psi
+ (a_4 -6k_d b_2) \psi \nabla \cdot \nabla \psi \right]  \Bigg) \mathbf{I} \\
& + \dfrac{G \delta_x^4}{6} \left[ (a_1 - 6k_d b_1) \nabla \psi \nabla \psi
+ (a_2 -6k_d b_1) \psi \nabla \nabla \psi \right] + O(\nabla^4) .
\end{split}
\end{equation}
Similarly, a steady-state one-dimensional flat interface along $y$ direction
is considered to determine the coexistence densities. The corresponding
normal pressure $P_n$ is
\begin{equation} \label{eq-pn}
P_n =P_{xx} = \dfrac{\rho c^2}{3} + \dfrac{G \delta_x^2}{2} \psi^2 +
\dfrac{G \delta_x^4}{6} \left[ -\dfrac{1+12k_d}{2} \left( \dfrac{d\psi}{dx}
\right)^2 + (1-6k_d) \psi \dfrac{d^2\psi}{dx^2} \right],
\end{equation}
where Eqs. (\ref{eq-parameters-a03}) and (\ref{eq-parameters-b01}) have been
used for the simplification. After some algebra, the following mechanical
stability condition is obtained
\begin{equation} \label{eq-mechanical}
\int_{\rho_g^{}}^{\rho_l^{}} \left( p_0^{} - \dfrac{\rho c^2}{3} -
\dfrac{G \delta_x^2}{2} \psi^2 \right) \dfrac{\psi'}{\psi^{1+\epsilon}}
d\rho =0 \enskip \text{with} \enskip \epsilon = \dfrac{1 + 12k_d}{1 - 6k_d},
\end{equation}
and accordingly the coexistence densities ($\rho_g^{}$ and $\rho_l^{}$) can
be determined. From Eq. (\ref{eq-mechanical}), it can be seen that the free
parameters $a_{1-4}$ and $b_{1-2}$ make no difference to the coexistence
densities. Actually, the other macroscopic properties, including the density
profile across the phase interface and the surface tension, can also be
uniquely determined by the pressure tensor $\mathbf{P}$, even though there
exist the free parameters $a_{1-4}$ and $b_{1-2}$.
\par

\begin{figure}[htbp]
  \centering
  \includegraphics[scale=1,draft=\figdraft]{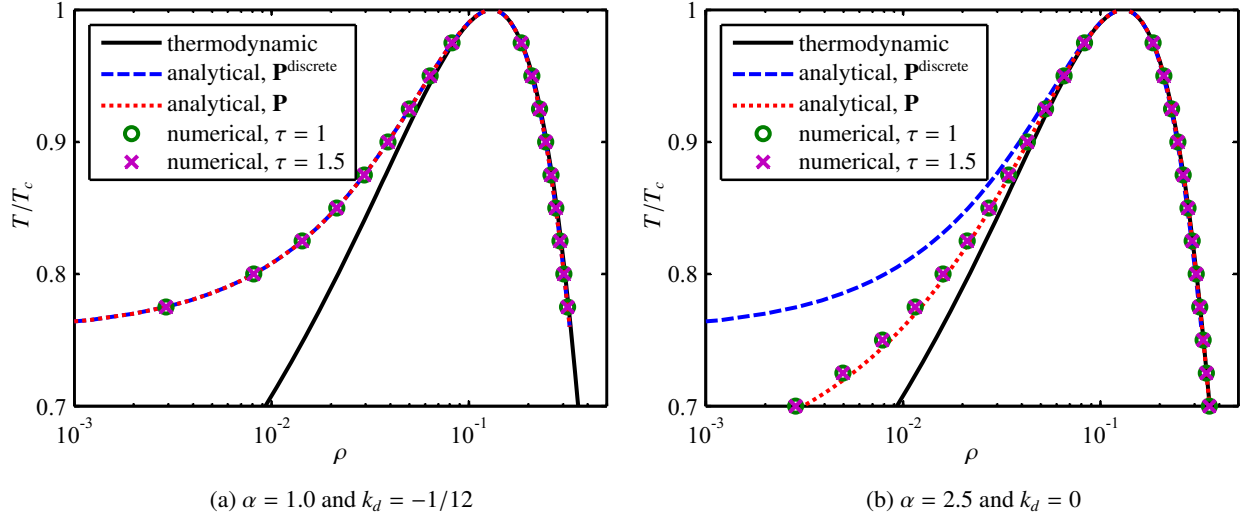}
  \caption{Comparisons of the coexistence curves given by the Maxwell
  construction (thermodynamic), the {\it discrete form} and {\it continuum
  form} pressure tensors ($\mathbf{P}^\text{discrete}$ and $\mathbf{P}$),
  and the numerical simulations ($\tau=1$ and $\tau=1.5$) for $\alpha=1.0$
  and $\alpha=2.5$.}
  \label{fig-coexistence-curve-comparisons}
\end{figure}

From the above analysis, we can see that the mechanical stability
conditions, which determine the coexistence densities, given by the two
forms of pressure tensors differ only in the parameter $\epsilon$ (see Eqs.
(\ref{eq-mechanical-discrete}) and (\ref{eq-mechanical})). For the {\it
discrete form} pressure tensor $\mathbf{P}^\text{discrete}$, $\epsilon=0$;
while for the {\it continuum form} pressure tensor $\mathbf{P}$, $\epsilon =
(1+12k_d) / (1-6k_d)$. To show the differences between
$\mathbf{P}^\text{discrete}$ and $\mathbf{P}$, the analytical coexistence
curves (coexistence densities versus temperature) are calculated by Eqs.
(\ref{eq-mechanical-discrete}) and (\ref{eq-mechanical}), respectively. For
comparisons, the thermodynamic results given by the Maxwell construction and
the numerical results given by the real simulation of a one-dimensional flat
interface are also presented. Here, the simulation is carried out on a $N_x
\times N_y = 1024 \times 8$ lattice with periodic boundary conditions in
both directions. The relaxation parameters are set as: $s_0^{} = s_j =1$,
$s_p = s_\varepsilon = 1/\tau$, $s_e = 1 / (5\tau-2)$ (i.e., $\sigma_e = 5
\sigma_p$), and $s_q = 1/ [0.5 + {\it \Lambda}/ (s_p^{-1} -0.5)]$ with ${\it
\Lambda} \equiv 1/12$. The density and velocity fields are initialized as
\begin{subequations} \label{eq-initialize-flat-interface}
\begin{equation}
\rho (\mathbf{x}) = \dfrac{\rho_g^\text{thermo} + \rho_l^\text{thermo}}{2} +
\dfrac{\rho_g^\text{thermo} - \rho_l^\text{thermo}}{2}
\tanh \dfrac{2(|x-x_c|-r_0)}{W},
\end{equation}
\begin{equation}
\mathbf{u} (\mathbf{x}) = \mathbf{0},
\end{equation}
\end{subequations}
where $\rho_g^\text{thermo}$ and $\rho_l^\text{thermo}$ denote the
thermodynamic coexistence gas and liquid densities given by the Maxwell
construction, $x_c = \tfrac{\delta_x}{2} N_x$, $W = 5 \delta_x$, and $r_0 =
\tfrac{\delta_x}{4} N_x$. Fig. \ref{fig-coexistence-curve-comparisons} gives
the comparisons of the coexistence curves obtained by different ways.
Obviously, the numerical results are $\tau \text{-independent}$, which can
also be easily known from the theoretical analysis (see Eqs.
(\ref{eq-mechanical-discrete}) and (\ref{eq-mechanical})). For $\alpha=1$
(the classical equilibrium moment), the coefficient $k_d = -1/12$, and the
parameter $\epsilon=0$ for both $\mathbf{P}^\text{discrete}$ and
$\mathbf{P}$. As it can be seen from Fig.
\ref{fig-coexistence-curve-comparisons}(a), the coexistence curves predicted
by $\mathbf{P}^\text{discrete}$ and $\mathbf{P}$ are identical and agree
well with the numerical results. Actually, if we set the free parameter $b_1
= -2a_1$, $\mathbf{P}$ given by Eq. (\ref{eq-p}) is identical to
$\mathbf{P}^\text{discrete}$ given by Eq. (\ref{eq-p-discrete-limit}) when
$k_d = -1/12$ (i.e., $\mathbf{R}_\text{iso} = \tfrac{1}{12} \delta_x^2
\nabla \cdot \nabla \mathbf{F}$). For $\alpha \neq 1$ (the general
equilibrium moment), $\alpha = 2.5$ is chosen as an example, and then the
coefficient $k_d =0$ (i.e., $\mathbf{R}_\text{iso} = \mathbf{0}$). Thus,
there have $\epsilon =0$ for $\mathbf{P}^\text{discrete}$ while $\epsilon
=1$ for $\mathbf{P}$. As it can be seen from Fig.
\ref{fig-coexistence-curve-comparisons}(b), the coexistence curve predicted
by $\mathbf{P}^\text{discrete}$ deviates the numerical results obviously,
while the coexistence curve predicted by $\mathbf{P}$ is still in good
agreement with the numerical results.
\par

From the above analysis and comparisons, we can conclude that accurate
pressure tensor can be definitely obtained in the {\it continuum form} when,
and only when, the third-order isotropic term is considered, and the
classical {\it discrete form} pressure tensor is accurate only when $k_d =
-1/12$ ($\alpha =1$ or $\sigma_e = \sigma_p$). For the general equilibrium
moment with $\alpha \neq 1$, the third-order isotropic term can be exploited
to adjust the coexistence densities (mechanical stability condition) simply
and directly (as indicated by Eq. (\ref{eq-mechanical}) and illustrated by
Fig. \ref{fig-coexistence-curve-comparisons}). However, this approach has a
direct effect on the bulk viscosity and may cause numerical instability when
$\alpha$ deviates strongly from the classical value $1.0$. Therefore, in
next section, a consistent scheme for third-order additional term will be
proposed to adjust the coexistence densities, as well as the surface tension
simultaneously and independently.
\par

\section{Scheme for third-order additional term} \label{sec-scheme}
\subsection{LB model with additional term} \label{subsec-scheme}
In the framework of the present third-order analysis, a consistent scheme is
proposed to introduce additional term into the recovered macroscopic
equation, which can be used to independently adjust the coexistence
densities (mechanical stability condition) and surface tension. The
additional term is devised to be recovered at the third-order, just like the
existing terms $\mathbf{R}_\text{iso}$ and $\mathbf{R}_\text{aniso}$, and
thus it makes no difference to the Navier-Stokes level (second-order)
macroscopic equation. To introduce such additional term, the collision step
in the moment space (i.e., Eq. (\ref{eq-collision})) is changed to
\begin{equation} \label{eq-collision-new}
\bar{\mathbf{m}}(\mathbf{x}, t) = \mathbf{m}(\mathbf{x}, t) - \mathbf{S}
\left[ \mathbf{m}(\mathbf{x}, t) - \mathbf{m}^\text{eq}(\mathbf{x}, t)
\right] + \delta_t \left( \mathbf{I} - \dfrac{\mathbf{S}}{2} \right)
\mathbf{F}_m(\mathbf{x}, t) + \mathbf{S Q}_m (\mathbf{x}, t),
\end{equation}
where $\mathbf{Q}_m (\mathbf{x}, t)$ is the discrete additional term in the
moment space. Inspired by the idea of Li and Luo \cite{Li2013-Tension},
$\mathbf{Q}_m (\mathbf{x}, t)$ can be chosen in the following form
\begin{equation} \label{eq-qm}
\mathbf{Q}_m = \Big( 0, \; Q_{m1}, \; Q_{m2}, \; 0, \; 0, \; 0, \; 0, \;
Q_{m7}, \; Q_{m8} \Big)^\text{T}.
\end{equation}
To determine $\mathbf{Q}_m$, systematic analysis is necessary and will be
carried out in next section. The streaming step is described by Eq.
(\ref{eq-streaming}). The equilibrium moment $\mathbf{m}^\text{eq}$, the
discrete force term $\mathbf{F}_m$, and the macroscopic variables are still
given by Eqs. (\ref{eq-meq}), (\ref{eq-fm}), and (\ref{eq-rho-u}),
respectively. Here, it is very interesting to note that the
exact-difference-method (EDM) forcing scheme \cite{Kupershtokh2009}, which
has attracted much attention in the pseudopotential LB community, can be
reformulated in the form of Eq. (\ref{eq-collision-new}), as presented in
\ref{app-a}.
\par

\subsection{Theoretical analysis} \label{subsec-analysis}
With the new collision step given by Eq. (\ref{eq-collision-new}), the
corresponding Taylor series expansion of the MRT LBE in the moment space
becomes
\begin{equation} \label{eq-3rd-taylor-lbe-new}
(\mathbf{I} \partial_t + \mathbf{D}) \mathbf{m} + \dfrac{\delta_t}{2}
(\mathbf{I} \partial_t + \mathbf{D})^2 \mathbf{m} + \dfrac{\delta_t^2}{6}
(\mathbf{I} \partial_t + \mathbf{D})^3 \mathbf{m} + O(\delta_t^3) =
-\dfrac{\mathbf{S}}{\delta_t} (\mathbf{m} - \mathbf{m}^\text{eq}) +
\left( \mathbf{I} - \dfrac{\mathbf{S}}{2} \right) \mathbf{F}_m +
\dfrac{\mathbf{S}}{\delta_t} \mathbf{Q}_m.
\end{equation}
In order to make the additional term recovered at the third-order
($\varepsilon^3$), $\mathbf{Q}_m$ is assumed to be at the order of
$\varepsilon^2$ , i.e., $\mathbf{Q}_m = \varepsilon^2 \mathbf{Q}_m^{(2)}$.
Then, Eq. (\ref{eq-3rd-taylor-lbe-new}) can be rewritten in the consecutive
orders of $\varepsilon$ as follows
\begin{subequations} \label{eq-3rd-orders-lbe-new}
\begin{equation} \label{eq-3rd-0th-lbe-new}
\varepsilon^0: \; \mathbf{m}^{(0)} = \mathbf{m}^\text{eq},
\end{equation}
\begin{equation} \label{eq-3rd-1st-lbe-new}
\varepsilon^1: \; (\mathbf{I} \partial_{t1} + \mathbf{D}_1) \mathbf{m}^{(0)}-
\mathbf{F}_m^{(1)} = - \dfrac{\mathbf{S}}{\delta_t} \left( \mathbf{m}^{(1)}
+ \dfrac{\delta_t}{2} \mathbf{F}_m^{(1)} \right) ,
\end{equation}
\begin{equation} \label{eq-3rd-2nd-lbe-new}
\varepsilon^2: \; \partial_{t2} \mathbf{m}^{(0)} + (\mathbf{I} \partial_{t1}
+ \mathbf{D}_1) \mathbf{m}^{(1)} + \dfrac{\delta_t}{2} (\mathbf{I}
\partial_{t1} + \mathbf{D}_1)^2 \mathbf{m}^{(0)} =
-\dfrac{\mathbf{S}}{\delta_t} \mathbf{m}^{(2)}
+\dfrac{\mathbf{S}}{\delta_t} \mathbf{Q}_m^{(2)},
\end{equation}
\begin{equation} \label{eq-3rd-3rd-lbe-new}
\varepsilon^3: \; \left( \begin{aligned}
& \partial_{t3} \mathbf{m}^{(0)} + \partial_{t2} \mathbf{m}^{(1)} +
(\mathbf{I} \partial_{t1} + \mathbf{D}_1) \mathbf{m}^{(2)} + \delta_t
(\mathbf{I} \partial_{t1} + \mathbf{D}_1) \partial_{t2} \mathbf{m}^{(0)} + \\
& \dfrac{\delta_t}{2} (\mathbf{I} \partial_{t1} + \mathbf{D}_1)^2
\mathbf{m}^{(1)} + \dfrac{\delta_t^2}{6} (\mathbf{I} \partial_{t1} +
\mathbf{D}_1)^3 \mathbf{m}^{(0)}
\end{aligned} \right)
= - \dfrac{\mathbf{S}}{\delta_t} \mathbf{m}^{(3)} .
\end{equation}
\end{subequations}
From Eq. (\ref{eq-3rd-orders-lbe-new}), we can see that $\mathbf{Q}_m^{(2)}$
appears in the second-order ($\varepsilon^2$) equation and will have an
effect on the following third-order ($\varepsilon^3$) equation. According to
the second-order Chapman-Enskog analysis in Section \ref{sec-2nd}, only the
equations for the conserved moments ($m_0$, $m_3$, and $m_5$) in the
second-order equation are involved to recover the Navier-Stokes level
macroscopic equation. Therefore, further considering $Q_{m0}^{(2)} =
Q_{m3}^{(2)} = Q_{m5}^{(2)} \equiv 0$ (see Eq. (\ref{eq-qm})),
$\mathbf{Q}_m^{(2)}$ in Eq. (\ref{eq-3rd-2nd-lbe-new}) truly makes no
difference to the Navier-Stokes level macroscopic equation, i.e., Eq.
(\ref{eq-2nd-orders-macro}) can still be recovered from Eqs.
(\ref{eq-3rd-0th-lbe-new}), (\ref{eq-3rd-1st-lbe-new}), and
(\ref{eq-3rd-2nd-lbe-new}).
\par

To identify the additional term introduced by $\mathbf{Q}_m$ at the
third-order, a steady and stationary situation can be considered, as
analyzed in Section \ref{sec-3rd}. Then, Eq. (\ref{eq-3rd-orders-lbe-new})
can be simplified as
\begin{subequations} \label{eq-3rd-orders-lbe-simple-new}
\begin{equation}
\varepsilon^0: \; \mathbf{m}^{(0)} = \mathbf{m}^\text{eq},
\end{equation}
\begin{equation}
\varepsilon^1: \; \partial_{t1} \mathbf{m}^{(0)} +
\mathbf{D}_1 \mathbf{m}^{(0)} - \mathbf{F}_m^{(1)} =
- \dfrac{\mathbf{S}}{\delta_t} \left( \mathbf{m}^{(1)} + \dfrac{\delta_t}{2}
\mathbf{F}_m^{(1)} \right),
\end{equation}
\begin{equation}
\varepsilon^2: \; \partial_{t2} \mathbf{m}^{(0)} - \delta_t \mathbf{D}_1
\left( \mathbf{S}^{-1} - \dfrac{\mathbf{I}}{2} \right)
\left( \mathbf{D}_1 \mathbf{m}^{(0)} - \mathbf{F}_m^{(1)} \right) =
- \dfrac{\mathbf{S}}{\delta_t} \mathbf{m}^{(2)}
+ \dfrac{\mathbf{S}}{\delta_t} \mathbf{Q}_m^{(2)},
\end{equation}
\begin{equation}
\varepsilon^3: \; \partial_{t3} \mathbf{m}^{(0)} + \delta_t^2 \left[
\mathbf{D}_1 \left( \mathbf{S}^{-1} - \dfrac{\mathbf{I}}{2} \right)
\mathbf{D}_1 \left( \mathbf{S}^{-1} - \dfrac{\mathbf{I}}{2} \right)
\left( \mathbf{D}_1 \mathbf{m}^{(0)} - \mathbf{F}_m^{(1)} \right) -
\dfrac{1}{12} \mathbf{D}_1^3 \mathbf{m}^{(0)} \right] +
\mathbf{D}_1 \mathbf{Q}_m^{(2)} =
- \dfrac{\mathbf{S}}{\delta_t} \mathbf{m}^{(3)}.
\end{equation}
\end{subequations}
After the same processes performed in Section \ref{sec-3rd}, the following
third-order macroscopic equation can be recovered
\begin{equation}
\begin{cases}
\partial_t \rho =0, \\
\partial_t (\rho \mathbf{u}) = -\nabla(\tfrac13 \rho c^2) + \mathbf{F}
+ \mathbf{R}_\text{iso} + \mathbf{R}_\text{aniso} + \mathbf{R}_Q,
\end{cases}
\end{equation}
where $\mathbf{R}_Q$ is the third-order additional term introduced by
$\mathbf{Q}_m$ that is expressed as
\begin{equation} \label{eq-ruq}
\mathbf{R}_Q = -c^2 \left[
\partial_x\left(\tfrac16 Q_{m1} + \tfrac12 Q_{m7}\right) + \partial_y Q_{m8},
\;
\partial_x Q_{m8} + \partial_y\left(\tfrac16 Q_{m1} - \tfrac12 Q_{m7}\right)
\right]^\text{T}.
\end{equation}
From Eq. (\ref{eq-ruq}), it can be seen that $Q_{m2}$ makes no difference to
the third-order additional term.
\par

With the consideration of the additional term $\mathbf{R}_Q$, the {\it
continuum form} pressure tensor (see Eq. (\ref{eq-p-definition})) is
redefined as
\begin{equation}
\nabla \cdot \mathbf{P} = \nabla ( \tfrac13 \rho c^2 ) - \mathbf{F}
- \mathbf{R}_\text{iso} - \mathbf{R}_Q.
\end{equation}
In order to independently adjust the mechanical stability condition
(coexistence densities) and surface tension, we take
\begin{equation} \label{eq-ruq-term}
\mathbf{R}_Q = -\nabla \cdot \left[ k_1 G \delta_x^4 \nabla \psi \nabla \psi
+ k_2 G \delta_x^4 ( \nabla \psi \cdot \nabla \psi ) \mathbf{I} \right],
\end{equation}
and subsequently, we can finally obtain the {\it continuum form} pressure
tensor as follows (see Section \ref{subsec-pressure})
\begin{equation} \label{eq-p-new}
\begin{split}
\mathbf{P} = \Bigg( \dfrac{\rho c^2}{3} + \dfrac{G \delta_x^2}{2} \psi^2
& + \dfrac{G \delta_x^4}{6} \left[ \left( a_3 - 6k_d b_2 + 6k_2 \right)
\nabla \psi \cdot \nabla \psi + \left( a_4  - 6k_d b_2 \right)
\psi \nabla \cdot \nabla \psi \right]  \Bigg) \mathbf{I} \\
& + \dfrac{G \delta_x^4}{6} \left[ \left( a_1 - 6k_d b_1 + 6k_1 \right)
\nabla \psi \nabla \psi + \left(a_2 -6k_d b_1\right) \psi \nabla \nabla \psi
\right] + O(\nabla^4) .
\end{split}
\end{equation}
Here, $k_1$ and $k_2$ are the adjustable parameters. As compared with Eq.
(\ref{eq-p}), the introduction of $\mathbf{R}_Q$ given by Eq.
(\ref{eq-ruq-term}) only changes the coefficients before the terms $\nabla
\psi \nabla \psi$ and $( \nabla \psi \cdot \nabla \psi ) \mathbf{I}$ in Eq.
(\ref{eq-p-new}). Comparing Eq. (\ref{eq-ruq-term}) with Eq. (\ref{eq-ruq}),
we can choose
\begin{equation} \label{eq-qm178-continuum}
\begin{split}
Q_{m1} &= 3\left( k_1 + 2k_2 \right) G \delta_x^4 \dfrac{
\partial_x \psi \partial_x \psi + \partial_y \psi \partial_y \psi}{c^2}, \\
Q_{m7} &= k_1 G \delta_x^4 \dfrac{\partial_x \psi \partial_x \psi
- \partial_y \psi \partial_y \psi}{c^2}, \\
Q_{m8} &= k_1 G \delta_x^4 \dfrac{\partial_x \psi \partial_y \psi}{c^2}.
\end{split}
\end{equation}
Eq. (\ref{eq-qm178-continuum}) is in the continuum form. In real
application, the gradient of $\psi$, $\nabla \psi = ( \partial_x\psi, \,
\partial_y\psi ) ^\text{T}$, needs to be calculated by an isotropic central
scheme (ICS) as follows
\begin{equation}
\nabla \psi \approx \dfrac{1}{\delta_x^2} \sum \limits_{i=1}^8
\omega(|\mathbf{e}_i \delta_t|^2) \psi (\mathbf{x}+\mathbf{e}_i\delta_t)
\mathbf{e}_i\delta_t = - \dfrac{\mathbf{F}}{G \delta_x^2 \psi},
\end{equation}
where the nearest-neighbor interaction force (i.e., Eq.
(\ref{eq-interaction-force})), as a finite-difference gradient operator, is
utilized to simplify the ICS. Therefore, $Q_{m1}$, $Q_{m7}$, and $Q_{m8}$
can be further written in a discrete form as
\begin{equation} \label{eq-qm178-discrete}
\begin{split}
Q_{m1} &= 3\left(k_1+2k_2\right) \dfrac{|\mathbf{F}|^2}{G \psi^2 c^2},\\
Q_{m7} &= k_1 \dfrac{F_x^2 - F_y^2}{G \psi^2 c^2},\\
Q_{m8} &= k_1 \dfrac{F_x F_y}{G \psi^2 c^2}.
\end{split}
\end{equation}
In the Chapman-Enskog analysis, $\mathbf{F}$ is at the order of
$\varepsilon$. According to Eq. (\ref{eq-qm178-discrete}), $\mathbf{Q}_m$ is
at the order of $\varepsilon^2$, which is consistent with the aforementioned
assumption, and $\mathbf{R}_Q = -\nabla \cdot \big( k_1 G^{-1} \psi^{-2}
\mathbf{FF} + k_2 G^{-1} \psi^{-2} |\mathbf{F}|^2 \mathbf{I} \big)$ is at
the order of $\varepsilon^3$, which is consistent with the fact that
$\mathbf{R}_Q$ is recovered at the third-order. This consistency is the
reason why we call the present scheme for additional term a consistent
scheme. However, in previous works \cite{Li2012, Hu2015, Zarghami2015},
similar third-order terms, like $\nabla \cdot (h \mathbf{FF})$ ($h$ is a
coefficient), are inconsistently recovered and analyzed at the second-order.
Note that $Q_{m2}$ in $\mathbf{Q}_m$ is still undetermined. Based on the
third-order analysis, $Q_{m2}$ can be chosen arbitrarily, and it is set as
$Q_{m2} = -Q_{m1}$ in the present work.
\par

To show the adjustments of the mechanical stability condition and surface
tension by $\mathbf{R}_Q$, a steady-state one-dimensional flat interface
along $y$ direction is considered again. The normal pressure $P_n$ and
tangential pressure $P_\tau$ given by Eq. (\ref{eq-p-new}) are
\begin{subequations}
\begin{equation}
P_n = P_{xx} = \dfrac{\rho c^2}{3} + \dfrac{G \delta_x^2}{2} \psi^2 +
\dfrac{G \delta_x^4}{6} \left[ -\dfrac{1+12k_d-12k_1-12k_2}{2}
\left( \dfrac{d\psi}{dx} \right)^2 + \left(1-6k_d\right)
\psi \dfrac{d^2\psi}{dx^2} \right],
\end{equation}
\begin{equation}
P_\tau = P_{yy} = \dfrac{\rho c^2}{3} + \dfrac{G \delta_x^2}{2} \psi^2 +
\dfrac{G \delta_x^4}{6} \left[ \left( a_3 -6k_d b_2 +6k_2 \right)
\left( \dfrac{d\psi}{dx} \right)^2 + \left( a_4 -6k_d b_2 \right)
\psi \dfrac{d^2\psi}{dx^2} \right],
\end{equation}
\end{subequations}
where Eqs. (\ref{eq-parameters-a03}) and (\ref{eq-parameters-b01}) have been
used for the simplifications. Then, the mechanical stability condition and
surface tension can be obtained as
\begin{equation} \label{eq-mechanical-new}
\int_{\rho_g^{}}^{\rho_l^{}} \left( p_0^{} - \dfrac{\rho c^2}{3} - \dfrac{G
\delta_x^2}{2} \psi^2 \right) \dfrac{\psi'}{\psi^{1+\epsilon}} d\rho =0
\enskip \text{with} \enskip \epsilon = \dfrac{1 + 12k_d - 12k_1 - 12k_2}{1 -
6k_d},
\end{equation}
\begin{equation} \label{eq-surface-tension-new}
\sigma = \int_{-\infty}^{+\infty} \left( P_n - P_\tau \right) dx =
-\dfrac{G \delta_x^4}{6} \left( 1-6k_1 \right) \int_{\rho_g^{}}^{\rho_l^{}}
\psi'^2 \sqrt{\varrho} \, d\rho,
\end{equation}
where $\psi' = d\psi / d\rho$ and $\varrho = (d\rho / dx)^2$. From Eqs.
(\ref{eq-mechanical-new}) and (\ref{eq-surface-tension-new}), we can clearly
see that the mechanical stability condition and surface tension can be
adjusted by $k_1 + k_2$ and $k_1$, respectively.
\par

\subsection{Numerical validations} \label{subsec-validations}
Numerical simulations are then carried out to validate the above theoretical
analysis of the present scheme for third-order additional term. The basic
simulation parameters are chosen the same as in Section \ref{sec-results}.
The rest simulation parameters are set as follows: $\alpha=1$, $s_0^{} = s_j
=1$, $s_p = s_e = s_\varepsilon = 1/\tau$, and $s_q = 1/ [0.5 + {\it
\Lambda} / (s_p^{-1} - 0.5)]$ with ${\it \Lambda} \equiv 1/12$. Then, there
have $k_d = -1/12$ and $\epsilon = -8(k_1 + k_2)$. Considering $\tau$ makes
invisible difference to the numerical results, it is chosen as $\tau=1.5$
here. Note that, though $\alpha=1$ is chosen which means
$\mathbf{R}_\text{aniso} = \mathbf{0}$, it is still recommended to set ${\it
\Lambda} \equiv 1/12$. This is because that when the surface tension is
adjusted by $k_1$, anisotropic term introduced by $\mathbf{Q}_m$ at the
fifth-order may be amplified and then needs to be considered. By setting
${\it \Lambda} \equiv 1/12$, this anisotropic term can be eliminated, just
like $\mathbf{R}_\text{aniso}$. A fifth-order heuristic analysis on this
point is given in \ref{app-b}. What is more, setting ${\it \Lambda} \equiv
1/12$ can help reduce the spurious current based on our numerical tests.
\par

\begin{figure}[htbp]
  \centering
  \includegraphics[scale=1,draft=\figdraft]{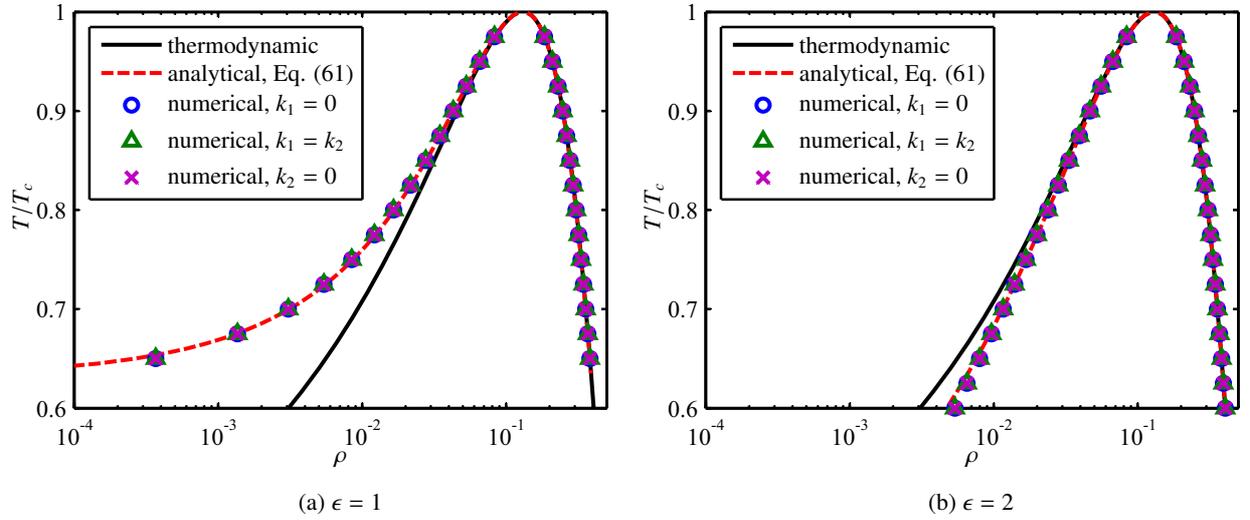}
  \caption{ Comparisons of the coexistence curves given by the Maxwell
  construction (thermodynamic), the mechanical stability condition
  (Eq. (\ref{eq-mechanical-new})), and the numerical simulations
  ($k_1=0$, $k_1 = k_2$, and $k_2=0$).}
  \label{fig-coexistence-curve-new}
\end{figure}

To validate the adjustment of the mechanical stability condition
(coexistence densities), the one-dimensional flat interface along $y$
direction is simulated on a $N_x \times N_y = 1024 \times 8$ lattice.
Periodic boundary conditions are applied in both directions and the initial
density and velocity fields are still given by Eq.
(\ref{eq-initialize-flat-interface}). The coexistence curves for the cases
$\epsilon=1$ and $\epsilon=2$ are shown in Fig.
\ref{fig-coexistence-curve-new}(a) and Fig.
\ref{fig-coexistence-curve-new}(b), respectively. It can be seen that the
numerical results are always in good agreement with the analytical results
predicted by the mechanical stability condition (i.e., Eq.
(\ref{eq-mechanical-new})), which validates the free adjustment of the
mechanical stability condition (coexistence densities) by the present scheme
and also verifies the theoretical analysis in Section \ref{subsec-analysis}.
What is more, Fig. \ref{fig-coexistence-curve-new} also shows that, as long
as $\epsilon = -8 (k_1 + k_2)$ keeps unvaried, the coexistence densities do
not vary with $k_1$. Thus, the surface tension can be independently adjusted
by varying the value of $k_1$ while fixing the value of $\epsilon$. Note
that, by properly setting the value of $\epsilon$, the coexistence densities
can be adjusted to approximate the thermodynamic results in real application
\cite{Li2012}.
\par

\begin{figure}[htbp]
  \centering
  \includegraphics[scale=1,draft=\figdraft]{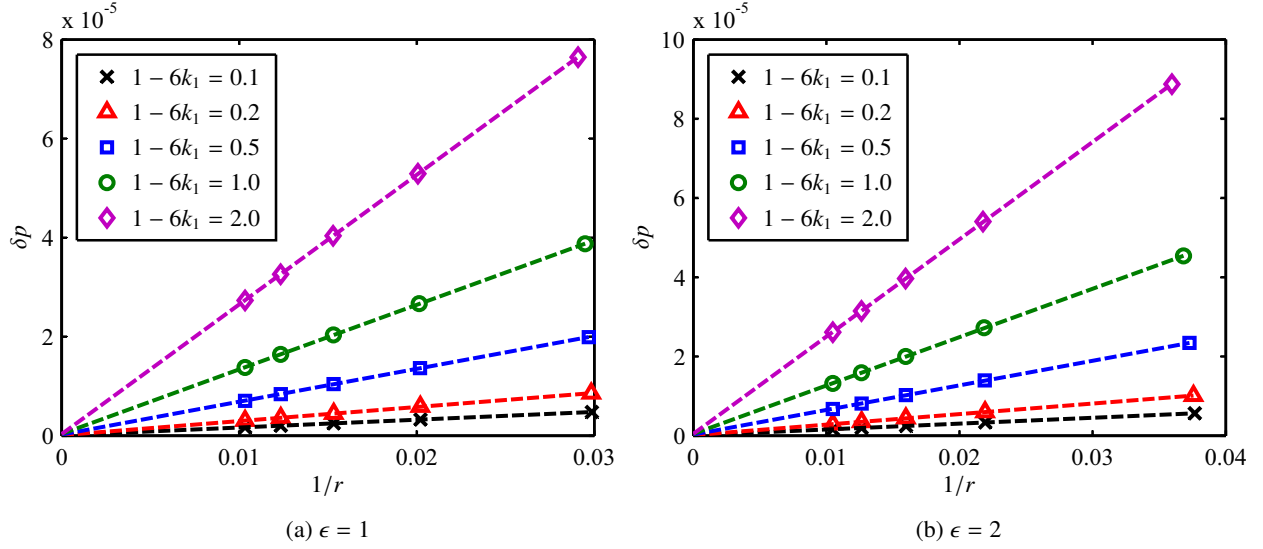}
  \caption{Variations of the pressure difference inside and outside of the
  droplet $\delta p$ with the reciprocal of the droplet radius $1/r$ for
  different $1-6k_1$. The dashed lines are the corresponding linear fits
  to the symbols.}
  \label{fig-surface-tension-new}
\end{figure}

To clearly show the adjustment of the surface tension, numerical simulations
of stationary droplets with different radii are carried out on a $N_x \times
N_y = 256 \times 265$ lattice with periodic boundary conditions in both
directions. The temperature is fixed at $T = 0.9 T_c$, and the initial
density and velocity fields are given by Eq. (\ref{eq-initialize-droplet})
except that the radius $r_0$ varies from $32 \delta_x$ to $96 \delta_x$. The
surface tension is numerically determined through the Laplace's law, i.e.,
$\delta p = p_\text{in}^{} - p_\text{out}^{} = \sigma / r$. Here,
$p_\text{in}^{}$ and $p_\text{out}^{}$ denote the pressure inside and
outside of the droplet, and $r$ is the final radius of the droplet. Fig.
\ref{fig-surface-tension-new} gives the numerical results of $\delta p$
versus $1/r$ for the cases $\epsilon=1$ and $\epsilon=2$ with $1-6k_1$
varying from $0.1$ to $2.0$. It clearly shows that the numerical results are
in good agreement with the linear fits denoted by the dashed lines, which
validates the Laplace's law. The slopes of the linear fits are equal to the
surface tensions, which are listed in Table \ref{table-surface-tension-new}.
As it can be seen, when $1-6k_1$ varies from $0.1$ to $2.0$, the surface
tension $\sigma$ varies from $1.5814 \times 10^{-4}$ to $2.6174 \times
10^{-3}$ for $\epsilon=1$ and from $1.4828 \times 10^{-4}$ to $2.4574 \times
10^{-3}$ for $\epsilon=2$. Note that, when the surface tension is too small,
it does not vary linearly with $1-6k_1$ as indicated by Eq.
(\ref{eq-surface-tension-new}), probably because that the influence of the
truncated higher-order terms on the surface tension is relatively strong
under this condition. What is more, when the surface tension is adjusted by
$1-6k_1$, the gas and liquid densities outside and inside the droplet vary
slightly though $\epsilon$ keeps unvaried, which can be seen from Table
\ref{table-surface-tension-new} for $r_0 = 64 \delta_x$ as an example. This
phenomenon is caused by the intrinsic property of the EOS, i.e., both the
gas and liquid phases are compressible to some degree.
\par

\begin{table}[htbp]
  \centering
  \caption{ Surface tensions ($\sigma$) determined through the Laplace's law
  for different $1-6k_1$, together with the gas and liquid densities
  ($\rho_g^{}$ and $\rho_l^{}$) given for $r_0=64\delta_x$.}
  \label{table-surface-tension-new}
  \tabcolsep=0em
  \newlength{\kb}\setlength{\kb}{1.38em plus 0.1em minus 0.1em}
  \begin{tabular}{c@{\hspace{0.5\kb}} c  c@{\hspace{\kb}}  c
                  c@{\hspace{\kb}}  c  c@{\hspace{\kb}}  c
                  c@{\hspace{\kb}}  c  c@{\hspace{\kb}}  c
                  c@{\hspace{\kb}}  c  c@{\hspace{0.5\kb}}  }
    \hline
    & \multirow{2}*{$1-6k_1$}
    && \multicolumn{5}{c}{$\epsilon=1$}
    && \multicolumn{5}{c}{$\epsilon=2$} & \\ \cline{4-8} \cline{10-14}
    &&&$\sigma$ && $\rho_g^{} (r_0=64\delta_x)$
    && $\rho_l^{} (r_0=64\delta_x)$
    && $\sigma$ && $\rho_g^{} (r_0=64\delta_x)$
    && $\rho_l^{} (r_0=64\delta_x)$ & \\
    \hline
    & 0.1 && $1.5814 \times 10^{-4}$ && $4.3726 \times 10^{-2}$
          && $2.4743 \times 10^{-1}$ && $1.4828 \times 10^{-4}$
          && $4.7530 \times 10^{-2}$ && $2.4883 \times 10^{-1}$ & \\
    & 0.2 && $2.8370 \times 10^{-4}$ && $4.3708 \times 10^{-2}$
          && $2.4745 \times 10^{-1}$ && $2.6579 \times 10^{-4}$
          && $4.7515 \times 10^{-2}$ && $2.4884 \times 10^{-1}$ & \\
    & 0.5 && $6.6303 \times 10^{-4}$ && $4.3655 \times 10^{-2}$
          && $2.4750 \times 10^{-1}$ && $6.2125 \times 10^{-4}$
          && $4.7469 \times 10^{-2}$ && $2.4890 \times 10^{-1}$ & \\
    & 1.0 && $1.3038 \times 10^{-3}$ && $4.3566 \times 10^{-2}$
          && $2.4758 \times 10^{-1}$ && $1.2219 \times 10^{-3}$
          && $4.7392 \times 10^{-2}$ && $2.4898 \times 10^{-1}$ & \\
    & 2.0 && $2.6174 \times 10^{-3}$ && $4.3388 \times 10^{-2}$
          && $2.4776 \times 10^{-1}$ && $2.4574 \times 10^{-3}$
          && $4.7239 \times 10^{-2}$ && $2.4916 \times 10^{-1}$ & \\
    \hline
  \end{tabular}
\end{table}

\section{Conclusions} \label{sec-conclu}
In this paper, we have performed a third-order Chapman-Enskog analysis of
the MRT pseudopotential LB model for multiphase flow for the first time. The
third-order leading terms on the interaction force are successfully
identified in the recovered macroscopic equation, and then some theoretical
aspects, which are still unclear or inconsistent in the pseudopotential LB
model, are discussed and clarified. Firstly, the isotropic property of the
LBE is investigated specifically. Numerical tests show that the third-order
anisotropic term recovered by the LBE needs to be eliminated for multiphase
flow, which means the isotropy of the LBE should be third-order at least in
the pseudopotential LB model. As indicated by the present third-order
analysis, this can be realized by adopting the classical equilibrium moment
or setting the so-called ``magic'' parameter to $1/12$. Then, the
determination of the pressure tensor, which is of crucial importance for
multiphase flow, is analyzed. It is shown that when and only when the
third-order isotropic term recovered by the LBE is considered, accurate {\it
continuum form} pressure tensor can be obtained from the recovered
macroscopic equation. By contrast, as also demonstrated by numerical tests,
the classical {\it discrete form} pressure tensor is accurate only when the
third-order isotropic term is a specific one. Finally, in the framework of
the present third-order analysis, a consistent scheme for third-order
additional term is proposed. By the present scheme, the coexistence
densities (mechanical stability condition) and surface tension can be
adjusted independently, which have been validated by the subsequent
numerical tests. In summary, by performing a third-order Chapman-Enskog
analysis, the theoretical foundations for the pseudopotential LB model are
further consolidated in this work. Simultaneously, the application of the
pseudopotential LB model can be extended by the present consistent scheme
for third-order additional term.

\section*{Acknowledgements}
This work was supported by the National Natural Science Foundation of China
through Grants No. 51536005 and No. 51376130, and the National Basic
Research Program of China (973 Program) through Grant No. 2012CB720404.

\appendix

\section{Reformulation of the EDM forcing scheme} \label{app-a}
The single-relaxation-time (SRT) LBE for the EDM forcing scheme is written
as \cite{Kupershtokh2009}
\begin{equation} \label{eq-lbe-edm-srt}
f_i (\mathbf{x} + \mathbf{e}_i\delta_t, t+\delta_t) = f_i (\mathbf{x}, t) -
\dfrac{1}{\tau} \left[ f_i (\mathbf{x}, t) - f_i^\text{eq} (\rho, \mathbf{v})
\right] + \left[ f_i^\text{eq} (\rho, \mathbf{v}+\delta\mathbf{v}) -
f_i^\text{eq} (\rho, \mathbf{v}) \right],
\end{equation}
where $\rho \mathbf{v} = \sum\nolimits_{i=0}^8 \mathbf{e}_i f_i$, $\delta
\mathbf{v} = \delta_t \mathbf{F} / \rho$, and $f_i^\text{eq} (\rho,
\mathbf{v})$ is the equilibrium distribution function. The macroscopic
density $\rho$ and velocity $\mathbf{u}$ are defined as
\begin{equation}
\rho = \sum\limits_{i=0}^{8} f_i, \qquad
\rho \mathbf{u} = \sum\limits_{i=0}^{8} \mathbf{e}_i f_i
+ \dfrac{\delta_t}{2} \mathbf{F} = \rho \left( \mathbf{v} +
\dfrac{ \delta \mathbf{v} }{2} \right).
\end{equation}
The MRT LBE for the EDM forcing scheme can be easily extended from Eq.
(\ref{eq-lbe-edm-srt}). The corresponding collision step is
\begin{equation} \label{eq-collision-edm}
\bar{\mathbf{m}}(\mathbf{x}, t) = \mathbf{m}(\mathbf{x}, t) - \mathbf{S}
\left[\mathbf{m}(\mathbf{x}, t)-\mathbf{m}^\text{eq}(\rho, \mathbf{v})\right]
+ \left[ \mathbf{m}^\text{eq}(\rho, \mathbf{v}+\delta\mathbf{v})
- \mathbf{m}^\text{eq}(\rho, \mathbf{v}) \right],
\end{equation}
where $\mathbf{m}^\text{eq} (\rho, \mathbf{v}) = \mathbf{M} \big[
f_0^\text{eq} (\rho, \mathbf{v}), \, \cdots, \, f_8^\text{eq} (\rho,
\mathbf{v}) \big]^\text{T}$ is the equilibrium moment that can be given as
\begin{equation}
\mathbf{m}^\text{eq} (\rho, \mathbf{v}) = \left( \rho, \; -2 \rho + 3 \rho
\dfrac{|\mathbf{v}|^2}{c^2}, \; \alpha \rho - 3 \rho
\dfrac{|\mathbf{v}|^2}{c^2}, \; \rho \dfrac{v_x}{c}, \; -\rho
\dfrac{v_x}{c}, \; \rho \dfrac{v_y}{c}, \; -\rho \dfrac{v_y}{c}, \; \rho
\dfrac{v_x^2-v_y^2}{c^2}, \; \rho \dfrac{v_xv_y}{c^2} \right) ^\text{T}.
\end{equation}
Substituting the relations $\mathbf{u} = \mathbf{v} + \delta \mathbf{v} / 2$
and $\delta \mathbf{v} = \delta_t \mathbf{F} / \rho$ into Eq.
(\ref{eq-collision-edm}), Eq. (\ref{eq-collision-edm}) can be reformulated
as
\begin{equation} \label{eq-collision-edm-reformulate}
\begin{split}
\bar{\mathbf{m}}(\mathbf{x}, t) & = \mathbf{m}(\mathbf{x}, t) - \mathbf{S}
\left[ \mathbf{m}(\mathbf{x}, t) - \mathbf{m}^\text{eq}(\rho,
\mathbf{u} - \tfrac{\delta_t}{2\rho} \mathbf{F}) \right] + \left[
\mathbf{m}^\text{eq}(\rho, \mathbf{u} + \tfrac{\delta_t}{2\rho} \mathbf{F}) -
\mathbf{m}^\text{eq}(\rho, \mathbf{u} - \tfrac{\delta_t}{2\rho} \mathbf{F})
\right] \\
& = \mathbf{m}(\mathbf{x}, t) - \mathbf{S} \left[ \mathbf{m}(\mathbf{x}, t)
- \mathbf{m}^\text{eq}(\rho, \mathbf{u}) \right] +
\left( \begin{aligned}
& \left( \mathbf{I} - \tfrac{\mathbf{S}}{2} \right) \left[
\mathbf{m}^\text{eq}(\rho, \mathbf{u} + \tfrac{\delta_t}{2\rho} \mathbf{F}) -
\mathbf{m}^\text{eq}(\rho, \mathbf{u} - \tfrac{\delta_t}{2\rho} \mathbf{F})
\right] + \\
& \tfrac{\mathbf{S}}{2} \left[
\mathbf{m}^\text{eq}(\rho, \mathbf{u} + \tfrac{\delta_t}{2\rho} \mathbf{F}) +
\mathbf{m}^\text{eq}(\rho, \mathbf{u} - \tfrac{\delta_t}{2\rho} \mathbf{F}) -
2 \mathbf{m}^\text{eq}(\rho, \mathbf{u}) \right]
\end{aligned} \right) \\
& = \mathbf{m}(\mathbf{x}, t) - \mathbf{S} \left[ \mathbf{m}(\mathbf{x}, t)
- \mathbf{m}^\text{eq}(\rho, \mathbf{u}) \right] + \delta_t \left( \mathbf{I}
- \dfrac{\mathbf{S}}{2} \right) \mathbf{F}_m (\mathbf{x}, t)
+ \mathbf{S} \mathbf{Q}_m^\text{EDM} (\mathbf{x}, t),
\end{split}
\end{equation}
where
\begin{subequations}
\begin{equation}
\begin{split}
\mathbf{F}_m (\mathbf{x}, t) & = \dfrac{1}{\delta_t} \left[
\mathbf{m}^\text{eq}(\rho, \mathbf{u} + \tfrac{\delta_t}{2\rho} \mathbf{F}) -
\mathbf{m}^\text{eq}(\rho, \mathbf{u} - \tfrac{\delta_t}{2\rho} \mathbf{F})
\right] \\
& =  \left( 0, \; 6\dfrac{\mathbf{F}\cdot\mathbf{u}}{c^2}, \;
-6\dfrac{\mathbf{F}\cdot\mathbf{u}}{c^2}, \;
\dfrac{F_x}{c}, \; -\dfrac{F_x}{c}, \; \dfrac{F_y}{c}, \; -\dfrac{F_y}{c}, \;
2\dfrac{F_xu_x-F_yu_y}{c^2}, \; \dfrac{F_xu_y+F_yu_x}{c^2} \right) ^\text{T},
\end{split}
\end{equation}
\begin{equation}
\begin{split}
\mathbf{Q}_m^\text{EDM} (\mathbf{x}, t) & = \dfrac12 \left[
\mathbf{m}^\text{eq}(\rho, \mathbf{u} + \tfrac{\delta_t}{2\rho} \mathbf{F}) +
\mathbf{m}^\text{eq}(\rho, \mathbf{u} - \tfrac{\delta_t}{2\rho} \mathbf{F}) -
2 \mathbf{m}^\text{eq}(\rho, \mathbf{u}) \right] \\
& = \left( 0, \; \dfrac34 \dfrac{\delta_x^2 |\mathbf{F}|^2}{\rho c^4}, \;
-\dfrac34 \dfrac{\delta_x^2 |\mathbf{F}|^2}{\rho c^4}, \; 0, \; 0, \;
0, \; 0, \; \dfrac14 \dfrac{\delta_x^2 (F_x^2 - F_y^2) }{\rho c^4}, \;
\dfrac14 \dfrac{\delta_x^2 F_x F_y}{\rho c^4} \right) ^\text{T}.
\end{split}
\end{equation}
\end{subequations}
Obviously, Eqs. (\ref{eq-collision-edm-reformulate}) and
(\ref{eq-collision-new}) are the same except the different coefficients in
the discrete additional term $\mathbf{Q}_m$. Based on the above analysis,
the nature of the EDM forcing scheme is revealed from a new perspective.
\par

\section{Fifth-order heuristic analysis on $\mathbf{Q}_m$} \label{app-b}
Performing the Taylor series expansion of the streaming step (i.e., Eq.
(\ref{eq-streaming})) to fifth-order, and correspondingly the Taylor series
expansion of the MRT LBE in the moment space becomes
\begin{equation}
\left( \begin{aligned}
& (\mathbf{I} \partial_t + \mathbf{D}) \mathbf{m} + \dfrac{\delta_t}{2}
(\mathbf{I} \partial_t + \mathbf{D})^2 \mathbf{m} + \dfrac{\delta_t^2}{6}
(\mathbf{I} \partial_t + \mathbf{D})^3 \mathbf{m} + \\
& \dfrac{\delta_t^3}{24} (\mathbf{I} \partial_t + \mathbf{D})^4 \mathbf{m}
+ \dfrac{\delta_t^4}{120} (\mathbf{I} \partial_t + \mathbf{D})^5 \mathbf{m}
+ O(\delta_t^5)
\end{aligned} \right)
= -\dfrac{\mathbf{S}}{\delta_t} (\mathbf{m} - \mathbf{m}^\text{eq}) +
\left( \mathbf{I} - \dfrac{\mathbf{S}}{2} \right) \mathbf{F}_m +
\dfrac{\mathbf{S}}{\delta_t} \mathbf{Q}_m,
\end{equation}
which can be rewritten in the consecutive orders of $\varepsilon$ as
\begin{subequations}
\begin{equation}
\varepsilon^0: \; \mathbf{m}^{(0)} = \mathbf{m}^\text{eq},
\end{equation}
\begin{equation}
\varepsilon^1: \; (\mathbf{I} \partial_{t1} + \mathbf{D}_1) \mathbf{m}^{(0)}-
\mathbf{F}_m^{(1)} = - \dfrac{\mathbf{S}}{\delta_t} \left( \mathbf{m}^{(1)}
+ \dfrac{\delta_t}{2} \mathbf{F}_m^{(1)} \right) ,
\end{equation}
\begin{equation}
\varepsilon^2: \; \partial_{t2} \mathbf{m}^{(0)} + (\mathbf{I} \partial_{t1}
+ \mathbf{D}_1) \mathbf{m}^{(1)} + \dfrac{\delta_t}{2} (\mathbf{I}
\partial_{t1} + \mathbf{D}_1)^2 \mathbf{m}^{(0)} =
-\dfrac{\mathbf{S}}{\delta_t} \mathbf{m}^{(2)}
+\dfrac{\mathbf{S}}{\delta_t} \mathbf{Q}_m^{(2)},
\end{equation}
\begin{equation}
\varepsilon^3: \; \left( \begin{aligned}
& \partial_{t3} \mathbf{m}^{(0)} +
\partial_{t2} \mathbf{m}^{(1)} + (\mathbf{I} \partial_{t1} + \mathbf{D}_1)
\mathbf{m}^{(2)} + \delta_t (\mathbf{I} \partial_{t1} + \mathbf{D}_1)
\partial_{t2} \mathbf{m}^{(0)} + \\
& \dfrac{\delta_t}{2} (\mathbf{I} \partial_{t1} + \mathbf{D}_1)^2
\mathbf{m}^{(1)} + \dfrac{\delta_t^2}{6}
(\mathbf{I} \partial_{t1} + \mathbf{D}_1)^3 \mathbf{m}^{(0)}
\end{aligned} \right) = - \dfrac{\mathbf{S}}{\delta_t} \mathbf{m}^{(3)} ,
\end{equation}
\begin{equation}
\varepsilon^4: \; \left( \begin{aligned}
& \partial_{t4} \mathbf{m}^{(0)} + \partial_{t3} \mathbf{m}^{(1)} +
\partial_{t2} \mathbf{m}^{(2)} + (\mathbf{I} \partial_{t1} + \mathbf{D}_1)
\mathbf{m}^{(3)} + \delta_t (\mathbf{I} \partial_{t1} + \mathbf{D}_1)
\partial_{t3} \mathbf{m}^{(0)} + \dfrac{\delta_t}{2} \partial_{t2}^2
\mathbf{m}^{(0)} + \\
& \delta_t (\mathbf{I} \partial_{t1} + \mathbf{D}_1) \partial_{t2}
\mathbf{m}^{(1)} + \dfrac{\delta_t}{2} (\mathbf{I} \partial_{t1} +
\mathbf{D}_1)^2 \mathbf{m}^{(2)} + \dfrac{\delta_t^2}{2} (\mathbf{I}
\partial_{t1} + \mathbf{D}_1)^2 \partial_{t2} \mathbf{m}^{(0)} + \\
& \dfrac{\delta_t^2}{6} (\mathbf{I} \partial_{t1} + \mathbf{D}_1 )^3
\mathbf{m}^{(1)} + \dfrac{\delta_t^3}{24} (\mathbf{I} \partial_{t1} +
\mathbf{D}_1)^4 \mathbf{m}^{(0)}
\end{aligned} \right) = - \dfrac{\mathbf{S}}{\delta_t} \mathbf{m}^{(4)} ,
\end{equation}
\begin{equation}
\varepsilon^5: \;  \left( \begin{aligned}
& \partial_{t5} \mathbf{m}^{(0)} + \partial_{t4} \mathbf{m}^{(1)}
+ \partial_{t3} \mathbf{m}^{(2)} + \partial_{t2} \mathbf{m}^{(3)} +
(\mathbf{I} \partial_{t1} + \mathbf{D}_1) \mathbf{m}^{(4)} +
\delta_t \partial_{t2} \partial_{t3} \mathbf{m}^{(0)} +
\dfrac{\delta_t}{2} \partial_{t2}^2 \mathbf{m}^{(1)} + \\
& \delta_t (\mathbf{I} \partial_{t1} + \mathbf{D}_1) \partial_{t4}
\mathbf{m}^{(0)} +  \delta_t (\mathbf{I} \partial_{t1} + \mathbf{D}_1)
\partial_{t3} \mathbf{m}^{(1)} + \delta_t (\mathbf{I} \partial_{t1} +
\mathbf{D}_1) \partial_{t2} \mathbf{m}^{(2)} + \\
& \dfrac{\delta_t}{2} (\mathbf{I} \partial_{t1} + \mathbf{D}_1)^2
\mathbf{m}^{(3)} + \dfrac{\delta_t^2}{2} (\mathbf{I} \partial_{t1} +
\mathbf{D}_1)^2 \partial_{t3} \mathbf{m}^{(0)} + \dfrac{\delta_t^2}{2}
(\mathbf{I} \partial_{t1} + \mathbf{D}_1) \partial_{t2}^2 \mathbf{m}^{(0)} + \\
& \dfrac{\delta_t^2}{2} (\mathbf{I} \partial_{t1} + \mathbf{D}_1)^2
\partial_{t2} \mathbf{m}^{(1)} + \dfrac{\delta_t^2}{6}
(\mathbf{I} \partial_{t1} + \mathbf{D}_1)^3 \mathbf{m}^{(2)} +
\dfrac{\delta_t^3}{6} (\mathbf{I} \partial_{t1} + \mathbf{D}_1)^3
\partial_{t2} \mathbf{m}^{(0)} + \\
& \dfrac{\delta_t^3}{24} (\mathbf{I} \partial_{t1} + \mathbf{D}_1)^4
\mathbf{m}^{(1)} + \dfrac{\delta_t^4}{120} (\mathbf{I} \partial_{t1} +
\mathbf{D}_1)^5 \mathbf{m}^{(0)}
\end{aligned} \right) = - \dfrac{\mathbf{S}}{\delta_t} \mathbf{m}^{(5)}.
\end{equation}
\end{subequations}
Similarly, a steady and stationary situation is considered, and the
lower-order equations are used to simplify the higher-order equations.
Finally, we can obtain
\begin{subequations} \label{eq-5th-orders-lbe-simple}
\begin{equation}  \label{eq-5th-0th-lbe-simple}
\varepsilon^0: \; \mathbf{m}^{(0)} = \mathbf{m}^\text{eq},
\end{equation}
\begin{equation}  \label{eq-5th-1st-lbe-simple}
\varepsilon^1: \; \partial_{t1} \mathbf{m}^{(0)} +
\mathbf{D}_1 \mathbf{m}^{(0)} - \mathbf{F}_m^{(1)} =
- \dfrac{\mathbf{S}}{\delta_t} \left( \mathbf{m}^{(1)} + \dfrac{\delta_t}{2}
\mathbf{F}_m^{(1)} \right),
\end{equation}
\begin{equation}  \label{eq-5th-2nd-lbe-simple}
\varepsilon^2: \; \partial_{t2} \mathbf{m}^{(0)} - \delta_t \mathbf{D}_1
\left( \mathbf{S}^{-1} - \tfrac{\mathbf{I}}{2} \right)
\left( \mathbf{D}_1 \mathbf{m}^{(0)} - \mathbf{F}_m^{(1)} \right) =
- \dfrac{\mathbf{S}}{\delta_t} \mathbf{m}^{(2)}
+ \dfrac{\mathbf{S}}{\delta_t} \mathbf{Q}_m^{(2)},
\end{equation}
\begin{equation}  \label{eq-5th-3rd-lbe-simple}
\varepsilon^3: \; \partial_{t3} \mathbf{m}^{(0)} +
\left( \begin{aligned}
& \delta_t^2 \left[ \mathbf{D}_1 \left( \mathbf{S}^{-1} -
\tfrac{\mathbf{I}}{2} \right) \mathbf{D}_1 \left( \mathbf{S}^{-1} -
\tfrac{\mathbf{I}}{2} \right) \left( \mathbf{D}_1 \mathbf{m}^{(0)} -
\mathbf{F}_m^{(1)} \right) - \tfrac{1}{12} \mathbf{D}_1^3 \mathbf{m}^{(0)}
\right] + \\
& \mathbf{D}_1 \mathbf{Q}_m^{(2)}
\end{aligned} \right)
= - \dfrac{\mathbf{S}}{\delta_t} \mathbf{m}^{(3)},
\end{equation}
\begin{equation}  \label{eq-5th-4th-lbe-simple}
\varepsilon^4: \; \partial_{t4} \mathbf{m}^{(0)} -
\left( \begin{aligned}
& \delta_t^3 \left[ \begin{aligned}
& \mathbf{D}_1 \left(\mathbf{S}^{-1} - \tfrac{\mathbf{I}}{2}
\right) \mathbf{D}_1 \left(\mathbf{S}^{-1} - \tfrac{\mathbf{I}}{2} \right)
\mathbf{D}_1 \left(\mathbf{S}^{-1} - \tfrac{\mathbf{I}}{2} \right)
\left( \mathbf{D}_1 \mathbf{m}^{(0)} - \mathbf{F}_m^{(1)} \right) + \\
& \tfrac{1}{12} \mathbf{D}_1^3 \left( \mathbf{S}^{-1} - \tfrac{\mathbf{I}}{2}
\right) \left( \mathbf{D}_1 \mathbf{m}^{(0)} - \mathbf{F}_m^{(1)} \right) +
\tfrac{1}{12} \mathbf{D}_1 \left( \mathbf{S}^{-1} - \tfrac{\mathbf{I}}{2}
\right) \mathbf{D}_1^3 \mathbf{m}^{(0)}
\end{aligned} \right] + \\
& \delta_t \mathbf{D}_1 \left( \mathbf{S}^{-1} - \tfrac{\mathbf{I}}{2}
\right) \mathbf{D}_1 \mathbf{Q}_m^{(2)}
\end{aligned} \right)
= - \dfrac{\mathbf{S}}{\delta_t} \mathbf{m}^{(4)},
\end{equation}
\begin{equation}  \label{eq-5th-5th-lbe-simple}
\varepsilon^5: \; \partial_{t5} \mathbf{m}^{(0)} +
\left( \begin{aligned}
& \delta_t^4 \left[ \begin{aligned}
& \mathbf{D}_1 \left(\mathbf{S}^{-1} - \tfrac{\mathbf{I}}{2} \right)
\mathbf{D}_1 \left(\mathbf{S}^{-1} - \tfrac{\mathbf{I}}{2} \right)
\mathbf{D}_1 \left(\mathbf{S}^{-1} - \tfrac{\mathbf{I}}{2} \right)
\mathbf{D}_1 \left(\mathbf{S}^{-1} - \tfrac{\mathbf{I}}{2} \right)
\left(\mathbf{D}_1 \mathbf{m}^{(0)} - \mathbf{F}_m^{(1)} \right) - \\
& \tfrac{1}{12} \mathbf{D}_1^3 \left(\mathbf{S}^{-1} -
\tfrac{\mathbf{I}}{2} \right) \mathbf{D}_1 \left(\mathbf{S}^{-1} -
\tfrac{\mathbf{I}}{2} \right) \left(\mathbf{D}_1 \mathbf{m}^{(0)} -
\mathbf{F}_m^{(1)} \right) - \\
& \tfrac{1}{12} \mathbf{D}_1 \left(\mathbf{S}^{-1} - \tfrac{\mathbf{I}}{2}
\right) \mathbf{D}_1^3 \left(\mathbf{S}^{-1} - \tfrac{\mathbf{I}}{2}
\right) \left(\mathbf{D}_1 \mathbf{m}^{(0)} - \mathbf{F}_m^{(1)} \right) - \\
& \tfrac{1}{12} \mathbf{D}_1 \left(\mathbf{S}^{-1} - \tfrac{\mathbf{I}}{2}
\right) \mathbf{D}_1 \left(\mathbf{S}^{-1} - \tfrac{\mathbf{I}}{2} \right)
\mathbf{D}_1^3 \mathbf{m}^{(0)} + \tfrac{1}{120} \mathbf{D}_1^5
\mathbf{m}^{(0)}
\end{aligned} \right] + \\
& \delta_t^2 \left[ \mathbf{D}_1 \left(\mathbf{S}^{-1} -
\tfrac{\mathbf{I}}{2} \right) \mathbf{D}_1 \left(\mathbf{S}^{-1} -
\tfrac{\mathbf{I}}{2} \right) \mathbf{D}_1 \mathbf{Q}_m^{(2)} -
\tfrac{1}{12} \mathbf{D}_1^3 \mathbf{Q}_m^{(2)} \right]
\end{aligned} \right)
= - \dfrac{\mathbf{S}}{\delta_t} \mathbf{m}^{(5)}.
\end{equation}
\end{subequations}
From Eq. (\ref{eq-5th-orders-lbe-simple}), we can see that the differential
operator before $\mathbf{Q}_m^{(2)}$ at the order of $\varepsilon^{n+2}$ is
the same as that before $\mathbf{m}^{(0)}$ at the order of $\varepsilon^n$.
For example, the differential operator before $\mathbf{Q}_m^{(2)}$ at the
fifth-order (see Eq. (\ref{eq-5th-5th-lbe-simple})) is $\mathbf{D}_1
(\mathbf{S}^{-1} - \mathbf{I}/2) \mathbf{D}_1 (\mathbf{S}^{-1} -
\mathbf{I}/2) \mathbf{D}_1 - \mathbf{D}_1^3 /12$, which is identical to the
differential operator before $\mathbf{m}^{(0)}$ at the third-order (see Eq.
(\ref{eq-5th-3rd-lbe-simple})). From the third-order analysis in Section
\ref{sec-3rd}, it is found that anisotropic term will appear at the
third-order if $\mathbf{m}^{(0)} = \mathbf{m}^\text{eq}$ is not chosen
specifically. Considering the form of $\mathbf{Q}_m$ given by Eqs.
(\ref{eq-qm}) and (\ref{eq-qm178-discrete}) does not coincide with the form
of $\mathbf{m}^\text{eq}$ no matter how $Q_{m2}$ is chosen, anisotropic term
about $\mathbf{Q}_m$ will appear at the fifth-order. Generally, the effect
of this fifth-order anisotropic term can be neglected. However, when the
surface tension is adjusted by $k_1$, this anisotropic term may be amplified
synchronously, and then needs to be considered. By setting the ``magic''
parameter ${\it \Lambda} \equiv 1/12$, this fifth-order anisotropic term can
be eliminated just as the third-order anisotropic term
$\mathbf{R}_\text{aniso}$, because of the same differential operator before
$\mathbf{Q}_m^{(2)}$ in Eq. (\ref{eq-5th-5th-lbe-simple}) and
$\mathbf{m}^{(0)}$ in Eq. (\ref{eq-5th-3rd-lbe-simple}).
\par

\section*{References}
\biboptions{sort&compress}

\end{document}